\documentclass[12pt,english]{article}

\usepackage[utf8]{inputenc}
\usepackage[T1]{fontenc}

\usepackage{amsmath,amssymb,amsfonts}
\usepackage{bm,bbm}
\usepackage{booktabs}
\usepackage{color}
\usepackage[dvips,letterpaper]{geometry}
\usepackage{epsfig}
\usepackage{fullpage}
\usepackage{graphicx,psfrag,epsf}
\usepackage{indentfirst}
\usepackage{lineno}
\usepackage{natbib}
\usepackage{setspace}
\usepackage{subfigure}
\usepackage{times}
\usepackage{url}
\usepackage{verbatim}

\usepackage{multirow}
\usepackage[table,xcdraw]{xcolor}
\usepackage{rotating}

 \vspace {-0.7cm}

\title{\bf The zero-adjusted log-symmetric quantile regression model applied to extramarital affairs data}

\author{
\normalsize
\textbf{Danúbia R. Cunha}$^{1}$\,, \textbf{Jose A. Divino}$^{1}$, \textbf{Helton Saulo}$^{2}$ \\[-0.05cm]
{\footnotesize $^{1}$Department of Economics, Universidade Cat\'olica de Bras\'{i}lia, Bras\'{i}lia, Brazil}\\[-0.05cm]
{\footnotesize $^{2}$Department of Statistics, Universidade de Bras\'{i}lia, Bras\'{i}lia, Brazil}\\[-0.05cm]
}

\date{}

\begin{document}

\maketitle

\vspace {-0.7cm}
\noindent{\bf Abstract.} \emph{In this work, we propose a zero-adjusted log-symmetric quantile regression model. Initially, we introduce zero-adjusted log-symmetric distributions, which allow for the accommodation of zeros. The estimation of the parameters is approached by the maximum likelihood method and a Monte Carlo simulation is performed to evaluate the estimates. Finally, we illustrate the proposed methodology with the use of a real extramarital affairs data set.}
%

\onehalfspacing

\section{Introduction}\label{sec:01}

The classical log-symmetric distributions (LS) are a generalization of the log-normal distribution and are particularly flexible in providing models with positive asymmetry and that have lighter/heavier tails than those of the log-normal distribution \citep{vanegasp:16a}. Regression models based on LS distributions were recently studied by \cite{vanegasp:15,vanegaspaula:17}, \cite{medeirosferrari:16}, in which problems such as semi-parametric approach, presence of censored data, or hypothesis tests, are investigated.

Recently, \cite{ssls:20} proposed a reparametrization of the LS distributions, here denoted by \textit{quantile}-LS distributions. The authors' idea was to insert a quantile parameter in the LS distributions and thus obtain the \textit{quantile}-LS distributions. Then, based on these distributions, the authors proposed parametric \textit{quantile}-LS quantile regression models. Parametric quantile regression models \citep{gilchrist:00} are an alternative to semi-parametric models (distribution-free) \citep{koenker:78}, and to models based on pseudo-likelihood using the asymmetric Laplace distribution or a mixture distribution. Based on \textit{quantile}-LS distributions, \cite{cds:21} proposed a quantile tobit model useful for modeling left censored data.

A limitation of the LS \citep{vanegasp:15} and \textit{quantile}-LS \citep{ssls:20} distributions, and consequently of their respective regression models, is the impossibility of modeling (without the need for any type of transformation) data sets that contain zeros, since such distributions have positive support. In this sense, a strategy to circumvent such a problem is to consider a mixture of a continuous distribution with support $(0,\infty)$ with a degenerate distribution with mass at zero; see, for example, \cite{aitchisonbrown:57} for the case of the log-normal distribution, \cite{Heller2006} for the inverse Gaussian distribution, \cite{Leiva2016} and \cite{Tomazella2019} for Birnbaum-Saunders distributions, and most recently \cite{cc:21,cosavalente:21} for LS distributions. The mixture of two components, distribution with support $(0,\infty)$ and a degenerate distribution with zero value, can be called ``zero adjusted'' as in \cite{Heller2006}, and it is similar to the \cite{cragg:71} approach.

In this context, the primary objective of this work is to propose a class of zero-adjusted log-symmetric quantile regression models. To this end, initially zero-adjusted \textit{quantile}-LS distributions are proposed, which are denoted by \textit{quantile}-ZALS distributions. Then, we propose the \textit{quantile}-ZALS regression models. The immediate advantage of the proposed methodology lies in the flexibility of the quantile approach, which allows considering the effects of explanatory variables over the spectrum of the dependent variable, in addition to the possibility of including zero, which is not possible in the quantile regression models studied by \cite{ssls:20}. The \textit{quantile}-ZALS regression models, proposed in this work, generalize for a quantile context, the works of \cite{aitchisonbrown:57}, \cite{cc:21,cosavalente:21} and \cite{Leiva2016}. Secondary objectives include: (i) to obtain the estimates of the model parameters of using the maximum likelihood method; (ii) to carry out a Monte Carlo simulation to assess the performance of the maximum likelihood estimates; and (iii) to illustrate the proposed methodology using a real data set on extramarital affairs.

The rest of this work proceeds as follows. In Section \ref{sec:2}, the \textit {quantile}-LS distributions introduced by \cite{ssls:20} are briefly discussed, then the \textit{quantile}-ZALS distributions are introduced. In Section \ref{sec:3}, the proposed \textit{quantile}-ZALS regression model is presented. In Section \ref{sec:4}, a Monte Carlo simulation is carried out to assess the performance of the maximum likelihood estimates. In Section \ref{sec:5}, an application to real data is performed. Finally, in Section \ref{sec:6} presents the final conclusions.

\section{\textit{Quantile}-LS and \textit{quantile}-ZALS distributions}\label{sec:2}

In this section, the \textit{quantile}-LS distributions introduced by \cite{ssls:20} are initially described. Then, a zero-adjusted version of these distributions, denoted by \textit{quantile}-ZALS, are proposed.

\subsection{\textit{Quantile}-LS distributions}\label{sec:2.1}

A random variable $T$ follows a \textit{quantile}-LS distribution if its probability density function (PDF) and cumulative distribution function (CDF) are given, respectively, by 
\begin{equation}\label{eq:quant:ft}
f_{T}(t;Q,\phi)
=
\dfrac{1}{\sqrt{\phi}\,t}\,
g\!\left(\frac{1}{\phi} \left[ \log(t)-\log(Q)+\sqrt{\phi}\,z_{p} \right]^2 \right), 
\quad 0<t<\infty,
\end{equation}
and
\begin{equation}\label{eq:quant:cd} 
 F_{T}(t;Q,\phi)
=
G\!\left(\frac{1}{\phi} \left[ \log(t)-\log(\lambda) \right]^2 \right)=
G\!\left(\frac{1}{\phi} \left[ \log(t)-\log(Q)+\sqrt{\phi}\,z_{p} \right]^2 \right),\quad 0<t<\infty,
\end{equation}
where $Q>0$ is a scale parameter and also the quantile of the distribution, $\phi>0$ is a power parameter, $g$ is a density generator, which may involve an extra parameter $\xi$, and $G(\omega)=\eta{\int^{\omega}_{-\infty} g(z^2)  \,\textrm{d}z }$ with $\omega\in\mathbb{R}$, with $\eta$ being a normalizing constant.  

\cite{ssls:20} have shown that if $T\sim\textrm{\textit{quantile}-LS} (Q,\phi,g)$, then the following properties hold: (P1) $cT\sim\textrm{\textit{quantile}-LS} (cQ,\phi,g)$, with $c>0$; (P2) $T^c\sim\textrm{\textit{quantile}-LS} (Q^c,c^2\phi,g)$, with $c>0$; and (P3) the quantile function is given by $Q_{T}(q;\lambda,\phi)=\lambda\exp\big(\sqrt{\phi}\,G^{-1}(q)\big), \quad q\in(0,1)$. The properties (P1) and (P2) imply that $T=Q\,\epsilon^{\sqrt{\phi}}$, where $\epsilon \sim \textrm{\textit{quantile}-LS} (1, 1, g)$.

The log-normal, Log-Student-$t$, log-power-exponential and extended Birnbaum-Saunders distributions are obtained as particular cases for $g$: 
\begin{itemize}
 \item Log-normal($Q,\phi$):  $ g(u)\propto\exp\left( -\frac{1}{2}u\right)$;   
 \item Log-Student-$t$($Q,\phi,\xi$):  $g(u) \propto \left(1+\frac{u}{\xi} \right)^{-\frac{\xi+1}{2}}$, $\xi>0$;    
\item Log-power-exponential($Q,\phi,\xi$):  $g(u) \propto \exp\left( -\frac{1}{2}u^{\frac{1}{1+\xi}}\right)$, $-1<{\xi}\leq{1}$;   
\item Extended Birnbaum-Saunders($Q,\phi,\xi$):  $g(u)\propto \cosh(u^{1/2})\exp\left(-\frac{2}{\xi^2}\sinh^2(u^{1/2}) \right) $, $\xi>0$. 
\end{itemize}

\subsection{\textit{Quantile}-ZALS distributions}\label{sec:2.2}

Consider a random variable $T\sim\textrm{\textit{quantile}-LS} (Q,\phi,g)$ with PDF and CDF given by \label{eq:quant:ft} and \label{eq:quant:cd}, respectively. We propose a \textit{quantile}-LS distribution that accommodates the zeros, denoted by \textit{quantile}-ZALS, by using a mixture approach given by
\begin{eqnarray}\label{eq:cragg}
\nonumber
g(z)&=&\pi\mathcal{I}_{(0)}(z)+(1-\pi)f_{T}(z)(1-\mathcal{I}_{(0)}(z)),\\\nonumber
 && \text{or}\\ 
g(z)&=&\pi^{\mathcal{I}_{(0)}(z)}\times\left\{(1-\pi)f_{T}(z)\right\}^{1-\mathcal{I}_{(0)}(z)}
\end{eqnarray}
where $0<\pi<1$ is a weight that determines the contribution of zeros, $f$ is the PDF of a random variable $leT$, and $\mathcal{I}_{A}(\cdot)$ is an indicator function, that is,
\begin{equation*}\label{eq:indic}
\mathcal{I}_{A}(x) =
\begin{cases}
1,      & \quad \text{if}\; x=A;\\
0      & \quad \text{if}\; x \neq A.
\end{cases}
\end{equation*}

The CDF given by Equation \eqref{eq:cragg} can be rewritten by replacing $f_{T}$ with \eqref{eq:quant:ft}, namely, 
\begin{eqnarray}\label{eq:cragg:pdf:zalogsym}
\nonumber
f_{Z}(z;Q,\phi,\pi)
&=&\pi\, \mathcal{I}_{\{0\}}(z) + 
\left\{ (1-\pi)\dfrac{1}{\sqrt{\phi}\,z}\,
g\!\left(\frac{1}{\phi} \left[ \log(z)-\log(Q)+\sqrt{\phi}\,z_{p} \right]^2 \right)\right\}
(1-\mathcal{I}_{(0)}(z)),\\ \nonumber
&&\quad\text{or}\\ 
f_{Z}(z;Q,\phi,\pi)
&=&\pi^{\mathcal{I}_{\{0\}}(z)} \times 
\left\{ (1-\pi)\dfrac{1}{\sqrt{\phi}\,z}\,
g\!\left(\frac{1}{\phi} \left[ \log(z)-\log(Q)+\sqrt{\phi}\,z_{p} \right]^2 \right)\right\}^{1-\mathcal{I}_{(0)}(z)}.
\end{eqnarray}
We use the notation $Z\sim\textrm{\textit{quantile}-ZALS}(Q,\phi,\pi,g)$. The CDF of $Z$ can be written as
\begin{equation*}\label{eq:cragg:cdf:zalogsym}
F_{Z}(z;Q,\phi,\pi) =
\begin{cases}
\pi,  & \quad \text{if}\; z=0,\\
\pi+(1-\pi)F_{T}(z;Q,\phi),      & \quad \text{if}\; z>0,
\end{cases}
\end{equation*}
where $F_{T}(\cdot)$ is the CDF of $T\sim\textrm{\textit{quantile}-LS} (Q,\phi,g)$ given in \eqref{eq:quant:cd}.

\section{\textit{Quantile}-ZALS regression model}\label{sec:3}

Based on the zero-adjusted quantile log-symmetric distributions proposed in Subsection~\ref{sec:2.2}, we propose the respective regression model. Consider $Z_1,\ldots,Z_n$ an independent random sample with $Z_i \sim \text{\textit{quantile}-ZALS}(Q_{i},\phi_{i},\pi_{i},g)$, for $i=1,\ldots,n$, such that
 \begin{equation*}\label{eq:cdf_zaqlogsym1}
F_{Z}(z_{i};Q_{i},\phi_{i},\pi_{i}) =
\begin{cases}
\pi_{i},  & \quad \text{if } z_{i}=0,\\
\pi_{i}+(1-\pi_{i})F_{T}(z_i;Q_{i},\phi_{i}),      & \quad \text{if } z_{i}>0,
\end{cases}
\end{equation*}
with
\begin{eqnarray*}\label{linkfunctions}\nonumber
Q_i    &=& \exp(\bm{x}_i^\top\bm \beta), \\ \nonumber
\phi_i &=& \exp(\bm{w}^{\top}_{i}\bm{\kappa}) \,\, \mbox{and} \\ 
\pi_i  &=& \Lambda(\bm{v}^{\top}_{i}\bm{\eta})=\frac{\exp(\bm{v}^{\top}_{i}\bm{\eta})}{1+\exp(\bm{v}^{\top}_{i}\bm{\eta})},
\end{eqnarray*}
where 
$\bm{\beta} =(\beta_0,\beta_1,\ldots,\beta_{k})^\top$, $\bm{\kappa}=(\kappa_0,\kappa_1,\ldots,{\kappa_{l}})^\top$ and 
$\bm{\eta}=(\eta_0,\eta_1,\ldots,{\eta_{m}})^\top$ are vectors of unknown parameters to be estimated, 
${\bm{x}}^{\top}_{i}= (1,x_{i1},\ldots, x_{ip})^\top$, 
${\bm{w}}^{\top}_{i} = (1,w_{i1}, \ldots, w_{iq})^\top$ and 
${\bm{v}}^{\top}_{i} = (1,v_{i1}, \ldots, v_{ir})^\top$
are the values of $p$, $q$ and $r$ explanatory variable associated with the quantile $Q_i$, relative dispersion $\phi_i$ and probability of drawing a zero $\pi_i$, respectively. Note that $\Lambda(x)=\frac{\exp(x)}{1+\exp(x)}$ is the logistic function.

The estimation of the parameters of the \textit{quantile}-ZALS regression model presented in \eqref{eq:cdf_zaqlogsym1} is performed using the maximum likelihood method. Let $Z_1,\ldots,\ldots,Z_n$ be an independent random sample such that $Z_i \sim \text{\textit{quantile}-ZALS}(Q_{i},\phi_{i},\pi_{i},g)$, and $z_1,\ldots,z_n$ be the corresponding observed values. Then, the likelihood function for $\boldsymbol{\theta}=(\bm{\beta}^{\top},\bm{\kappa}^{\top},\bm{\eta}^{\top})^{\top}$ can be written as
\begin{eqnarray}\label{eq:like}
\small
L(\boldsymbol{\theta})&=&\prod_{i=1}^{n}f_{Z}(z_i;Q_i,\phi_i,\pi_i)\\ \nonumber
&=& \underbrace{\prod_{i=1}^{n} \pi_i^{\mathcal{I}_{\{0\}}(z_i)}\, (1-\pi_i)^{1-\mathcal{I}_{\{0\}}(z_i)}}_
{L_1(\bm{\eta})} 
\underbrace{\prod_{i=1}^{n}
\left\{\dfrac{1}{\sqrt{\phi_i}\,z_i}\,
g\!\left(\frac{1}{\phi_i} \left[ \log(z_i)-\log(Q_i)+\sqrt{\phi_i}\,z_{p} \right]^2 \right)\right\}^{1-\mathcal{I}_{\{0\}}(z_i)}}_{L_2(\bm{\beta},\bm{\kappa})}.
\end{eqnarray}
Taking the logarithm in \ref{eq:like}, we obtain the log-likelihood function
\begin{eqnarray}\label{eq:loglike}
\small
\ell(\boldsymbol{\theta})&=&\sum_{i=1}^{n}\log(f_{Z}(z_i;Q_i,\phi_i,\pi_i))\\ \nonumber
&=& \underbrace{\sum_{i=1}^{n} {\mathcal{I}_{\{0\}}(z_i)}\log(\pi_i)\,+ (1-\mathcal{I}_{\{0\}}(z_i))\log(1-\pi_i)}_{\ell_1(\bm{\eta})} \\ \nonumber
&& \times \underbrace{\sum_{i=1}^{n}(1-\mathcal{I}_{\{0\}}(z_i))\log
\left\{\dfrac{1}{\sqrt{\phi_i}\,z_i}\,
g\!\left(\frac{1}{\phi_i} \left[ \log(z_i)-\log(Q_i)+\sqrt{\phi_i}\,z_{p} \right]^2 \right)\right\}}_{\ell_2(\bm{\beta},\bm{\kappa})}.
\end{eqnarray}
Note that in \ref{eq:loglike}, $\ell(\boldsymbol{\theta})$ is factored in two terms \citep{pacesalvan:97}, one that is associated with the probability of occurrence of zero, $\ell_1(\bm{\eta})$, and another that is associated with a continuous and positive part, $\ell_2(\bm{\beta},\bm{\kappa})$. Therefore, the maximum likelihood estimates can be obtained independently for $\boldsymbol{\theta}$ and $(\bm{\beta}^{\top},\bm{\kappa}^{\top})^{\top}$, that is, the maximization is performed separately for $\ell_1(\bm{\eta})$ and $\ell_2(\bm{\beta},\bm{\kappa})$. Nevertheless, as there is no analytical solution, an iterative procedure can be used for non-linear optimization, in particular the Broyden-Fletcher-Goldfarb-Shanno (BFGS) method is used in this work.

The \textit{quantile}-ZALS regression model proposed above can be interpreted as being divided into two equations:
\begin{itemize}
\item \textbf{Participation equation}
\begin{equation}\label{eq:indic}
\begin{cases}
\mathbb{P}(Z_i=0) = \pi_i= \frac{\exp(\bm{v}^{\top}_{i}\bm{\eta})}{1+\exp(\bm{v}^{\top}_{i}\bm{\eta})},    & \quad d_i=0,\,\text{if}\; z_i = 0,\\
\mathbb{P}(Z_i>0) = 1-\pi_i=\frac{1}{1+\exp(\bm{v}^{\top}_{i}\bm{\eta})} ,  & \quad d_i=1\,\text{if}\; z_i > 0,\\
\end{cases}
\end{equation}
where $d_i=1$ if the individual participates and $d_i=0$ otherwise.,
\item \textbf{Intensity equation}
\begin{equation}\label{eq:indic}
Q_i=Q(Z_i|d_i=1)=\exp(\bm{x}_i^\top\bm \beta),
\end{equation}
where $Q(Z_i|d_i=1)$ is the quantile o$f Z_i$ given that $d_i=1$.
\end{itemize}

\section{Monte Carlo simulation}\label{sec:4}

\begin{figure}[!ht]
\centering
\subfigure[$\widehat{\textrm{Bias}}$($\widehat{\beta}_i$)]{\includegraphics[height=5cm,width=5cm]{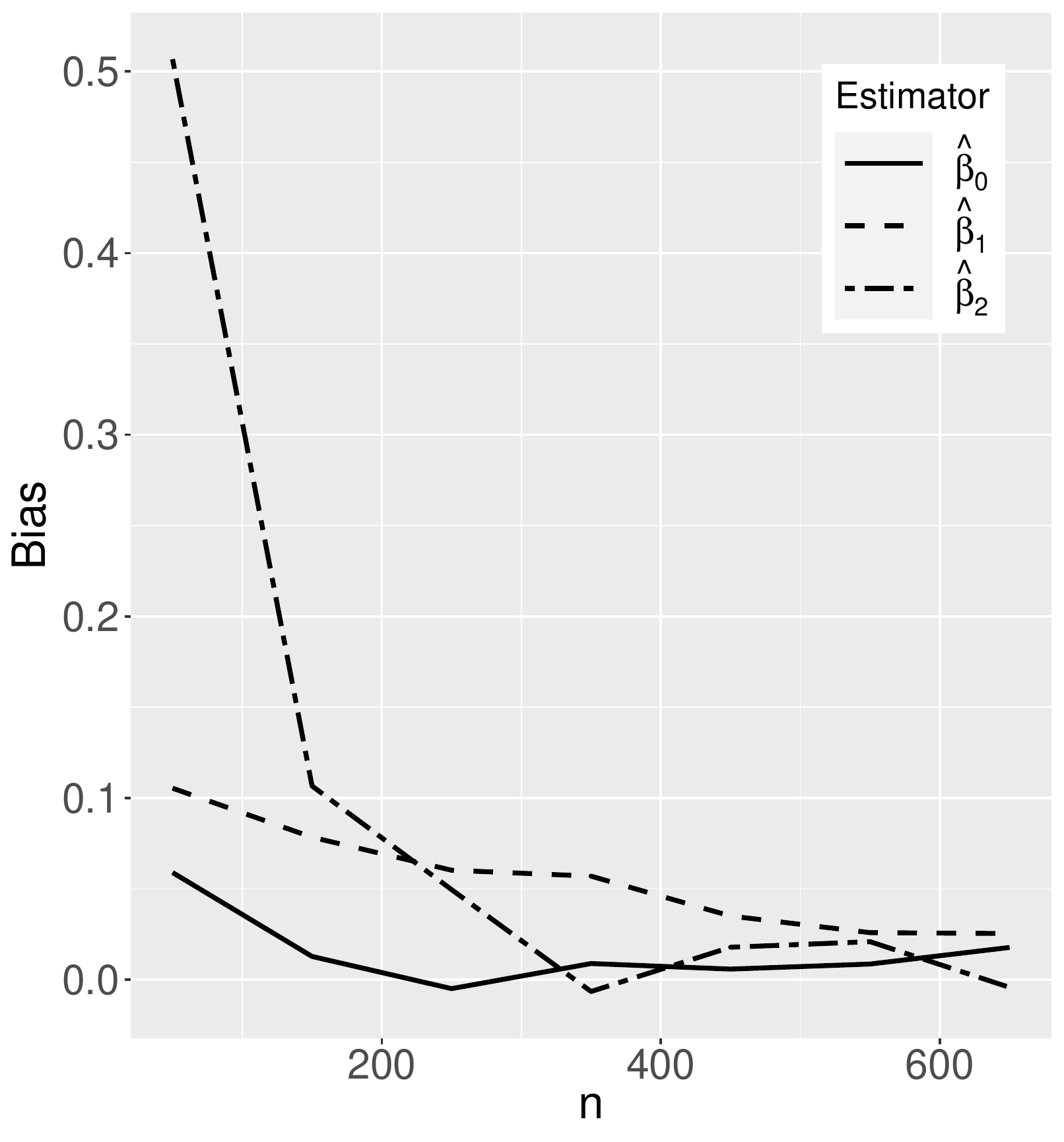}}
\subfigure[$\widehat{\textrm{MSE}}$($\widehat{\beta}_i$)]{\includegraphics[height=5cm,width=5cm]{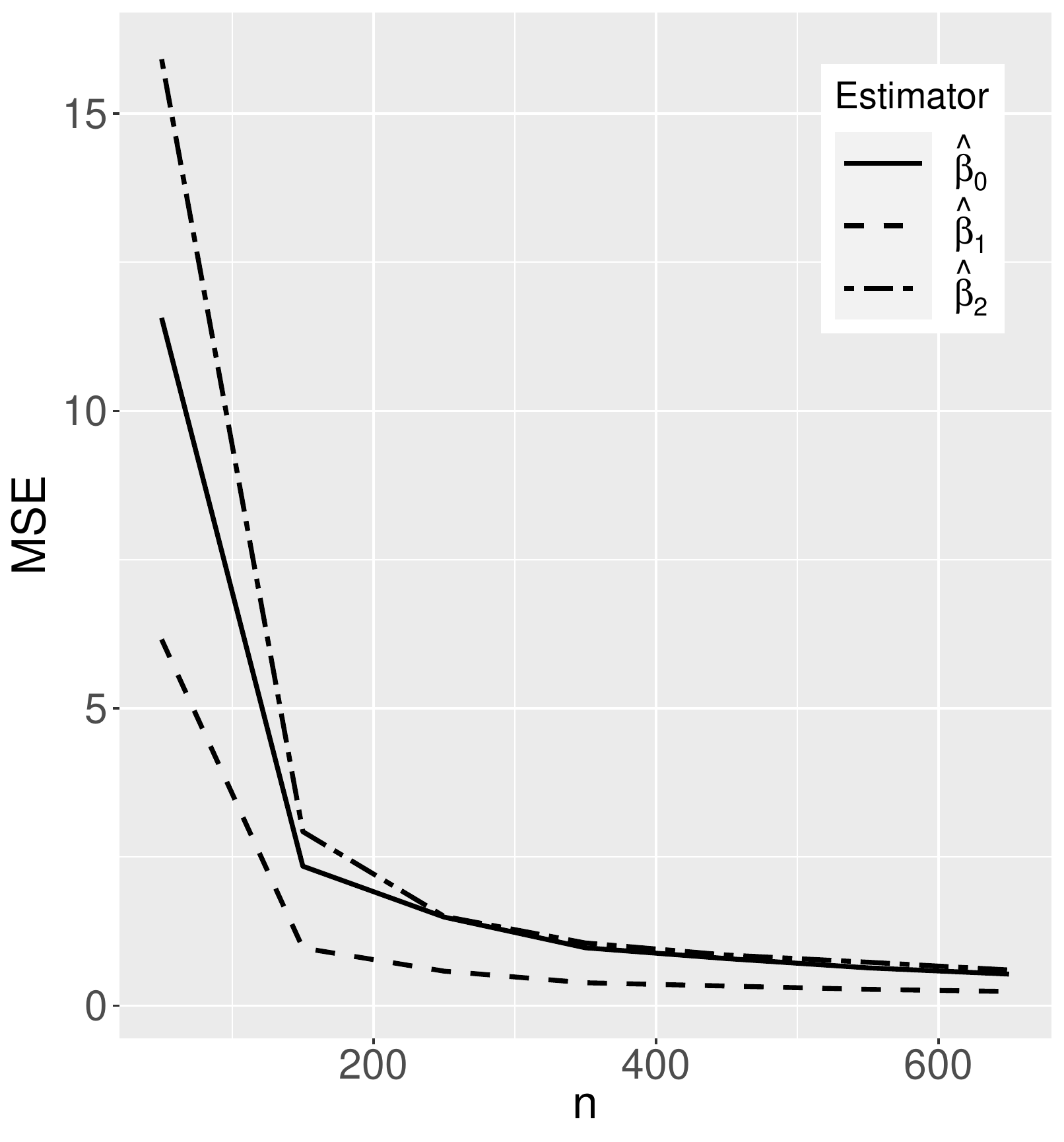}}
\subfigure[$\widehat{\textrm{Bias}}$($\widehat{\kappa}_i$)]{\includegraphics[height=5cm,width=5cm]{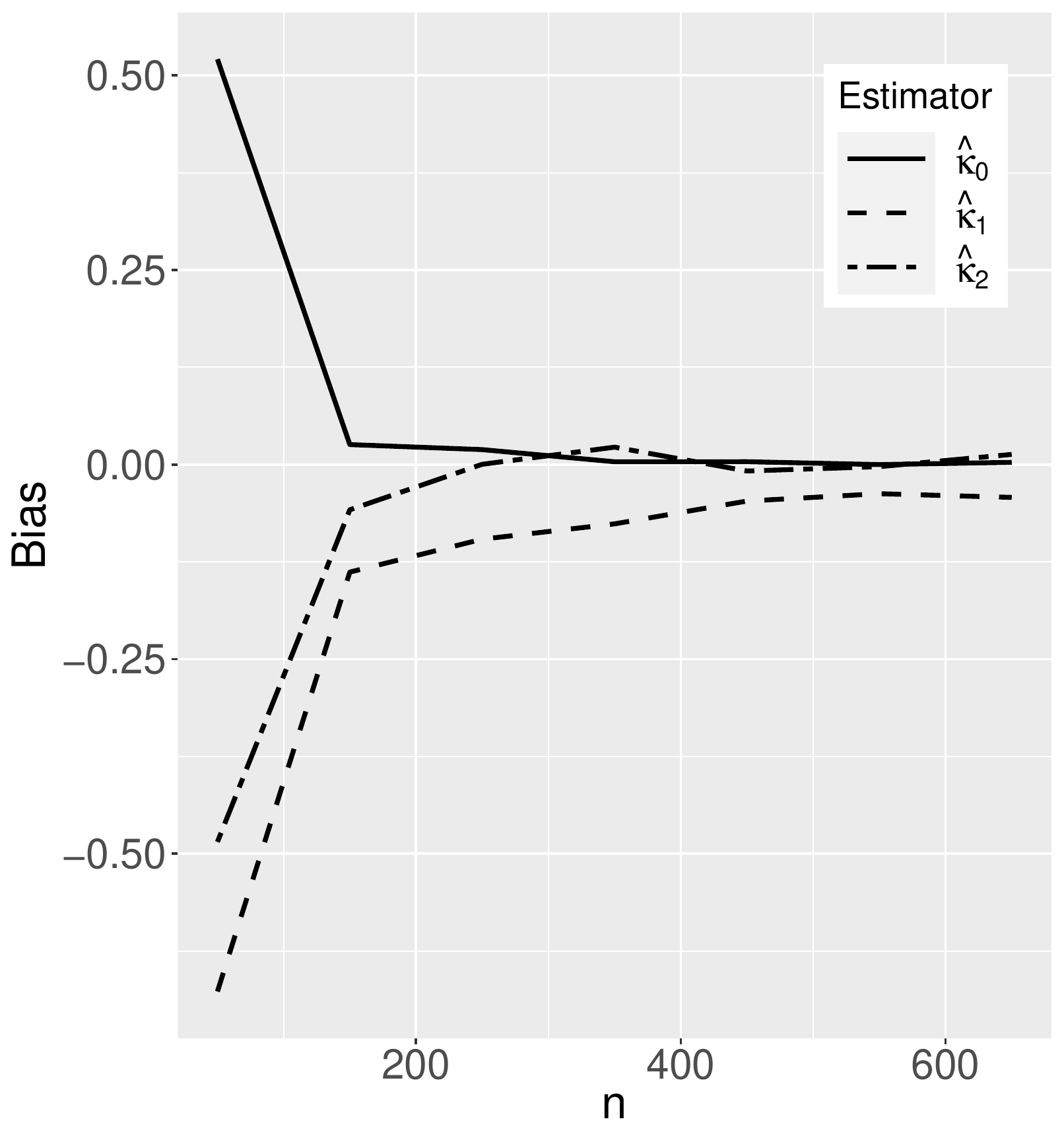}}
\subfigure[$\widehat{\textrm{MSE}}$($\widehat{\kappa}_i$)]{\includegraphics[height=5cm,width=5cm]{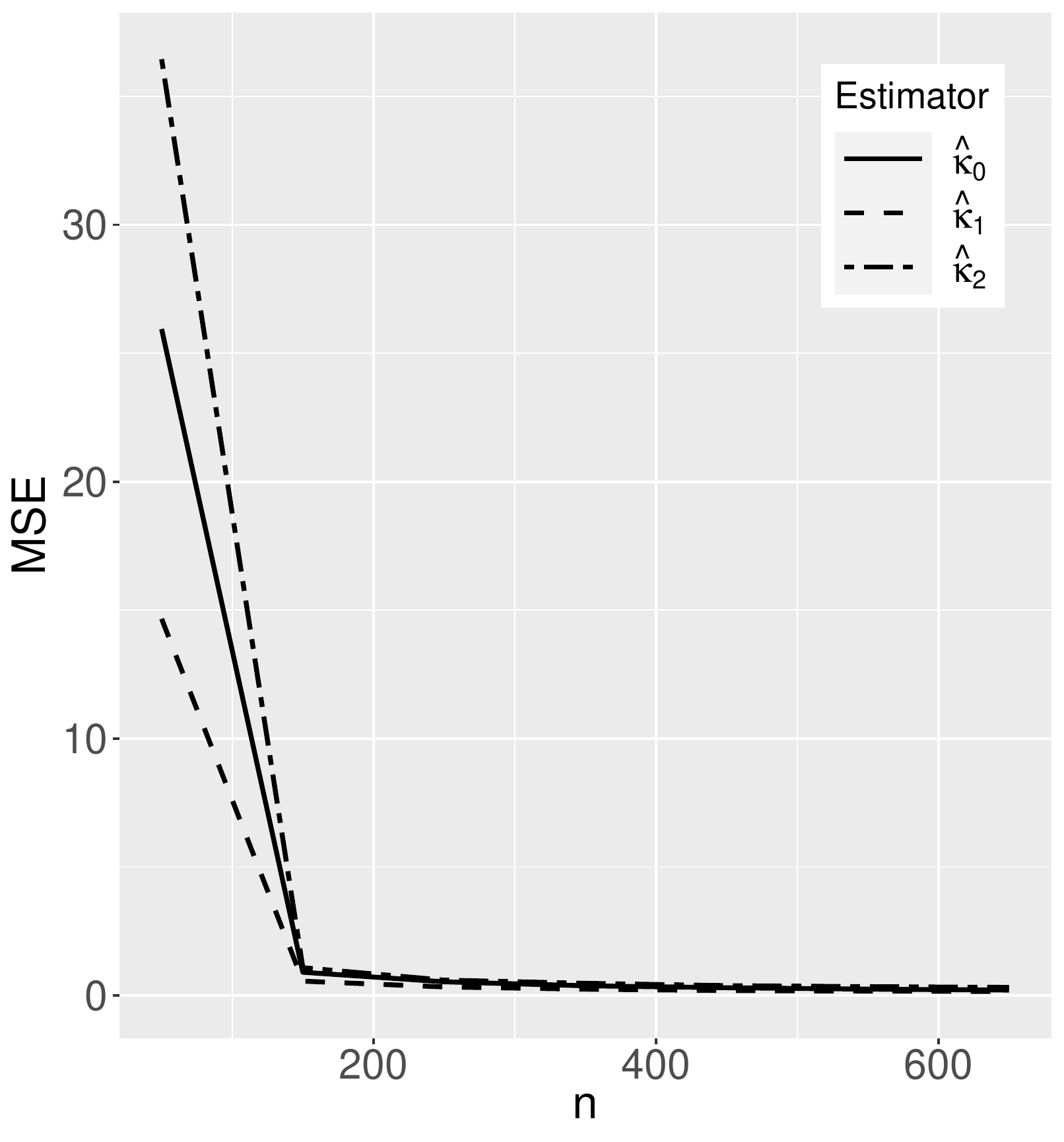}}
\subfigure[$\widehat{\textrm{Bias}}$($\widehat{\eta}_i$)]{\includegraphics[height=5cm,width=5cm]{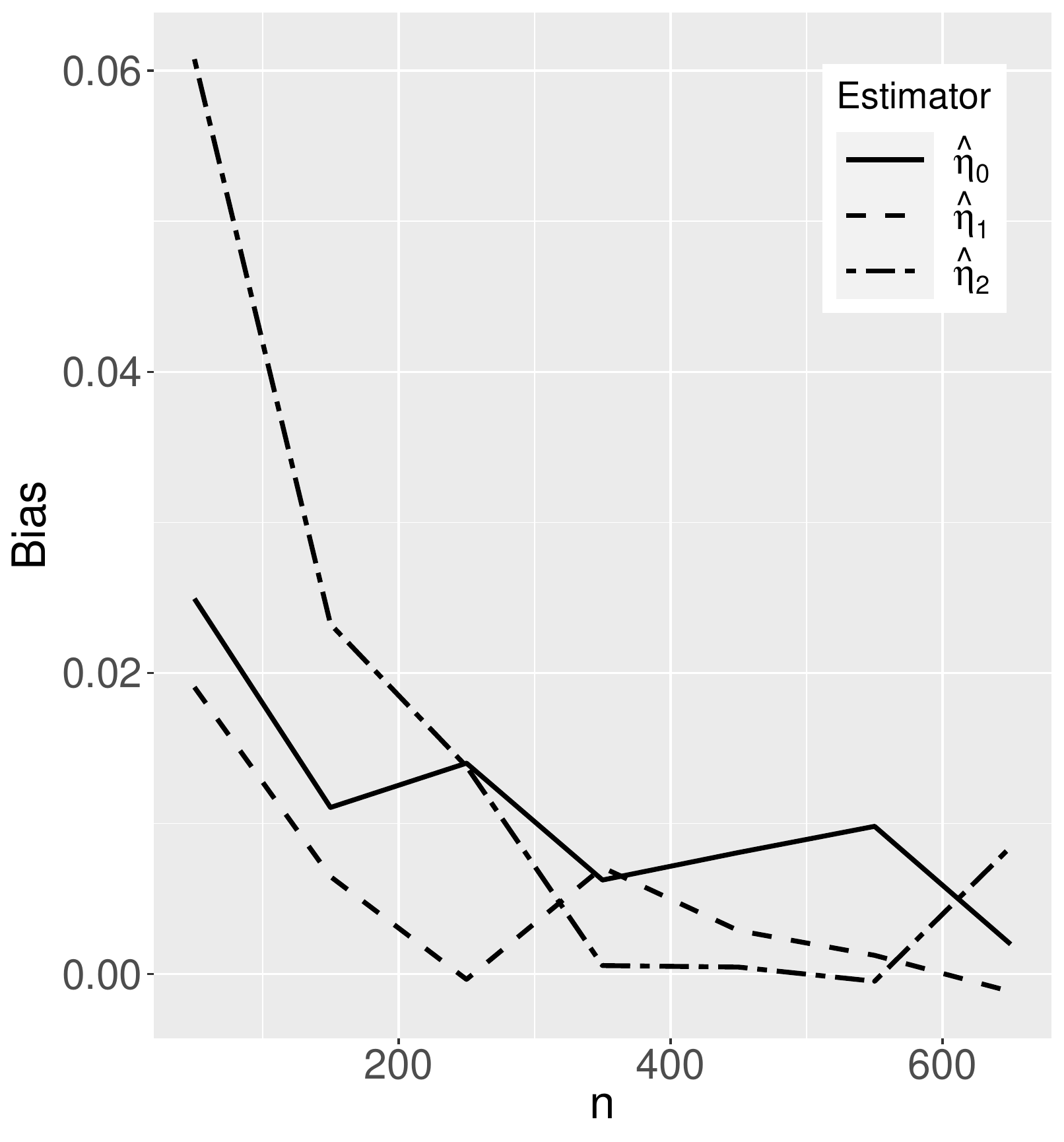}}
\subfigure[$\widehat{\textrm{MSE}}$($\widehat{\eta}_i$)]{\includegraphics[height=5cm,width=5cm]{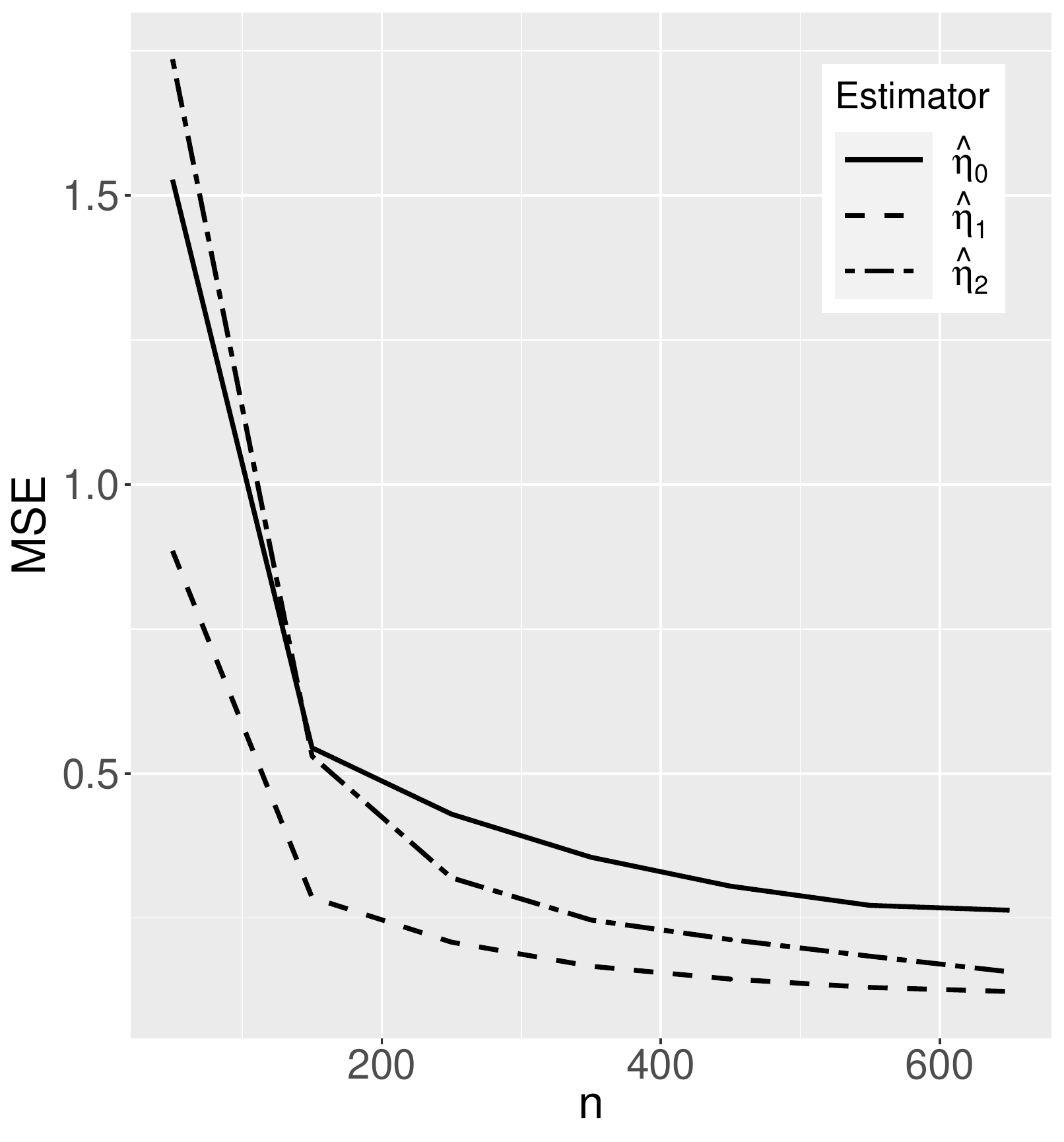}}
 \caption{\small {Bias and MSE estimates for $q=0.10$ ($i=\{0,1,2\}$).}}
\label{fig:sim1}
\end{figure}

\begin{figure}[!ht]
\centering
\subfigure[$\widehat{\textrm{Bias}}$($\widehat{\beta}_i$)]{\includegraphics[height=5cm,width=5cm]{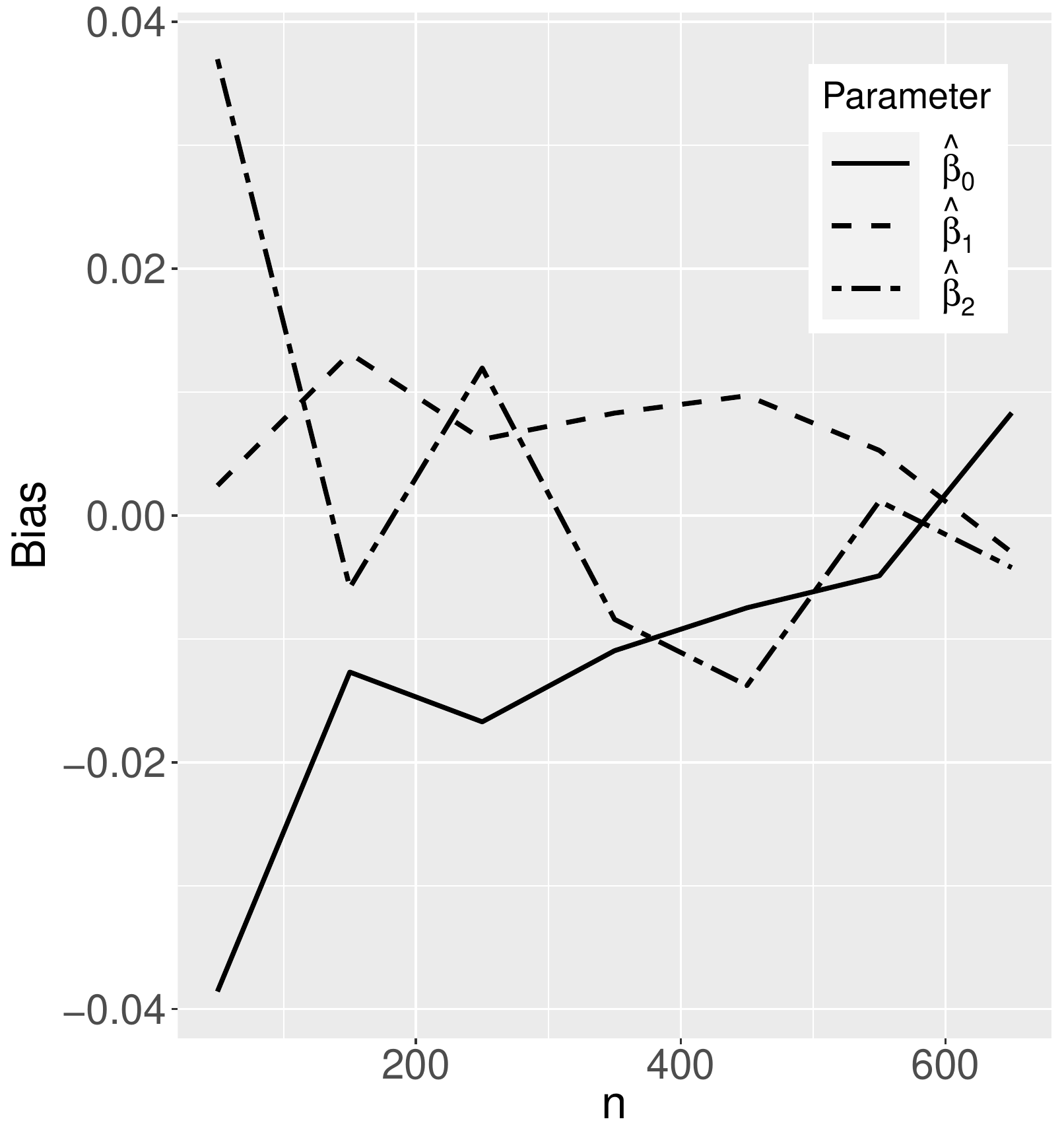}}
\subfigure[$\widehat{\textrm{MSE}}$($\widehat{\beta}_i$)]{\includegraphics[height=5cm,width=5cm]{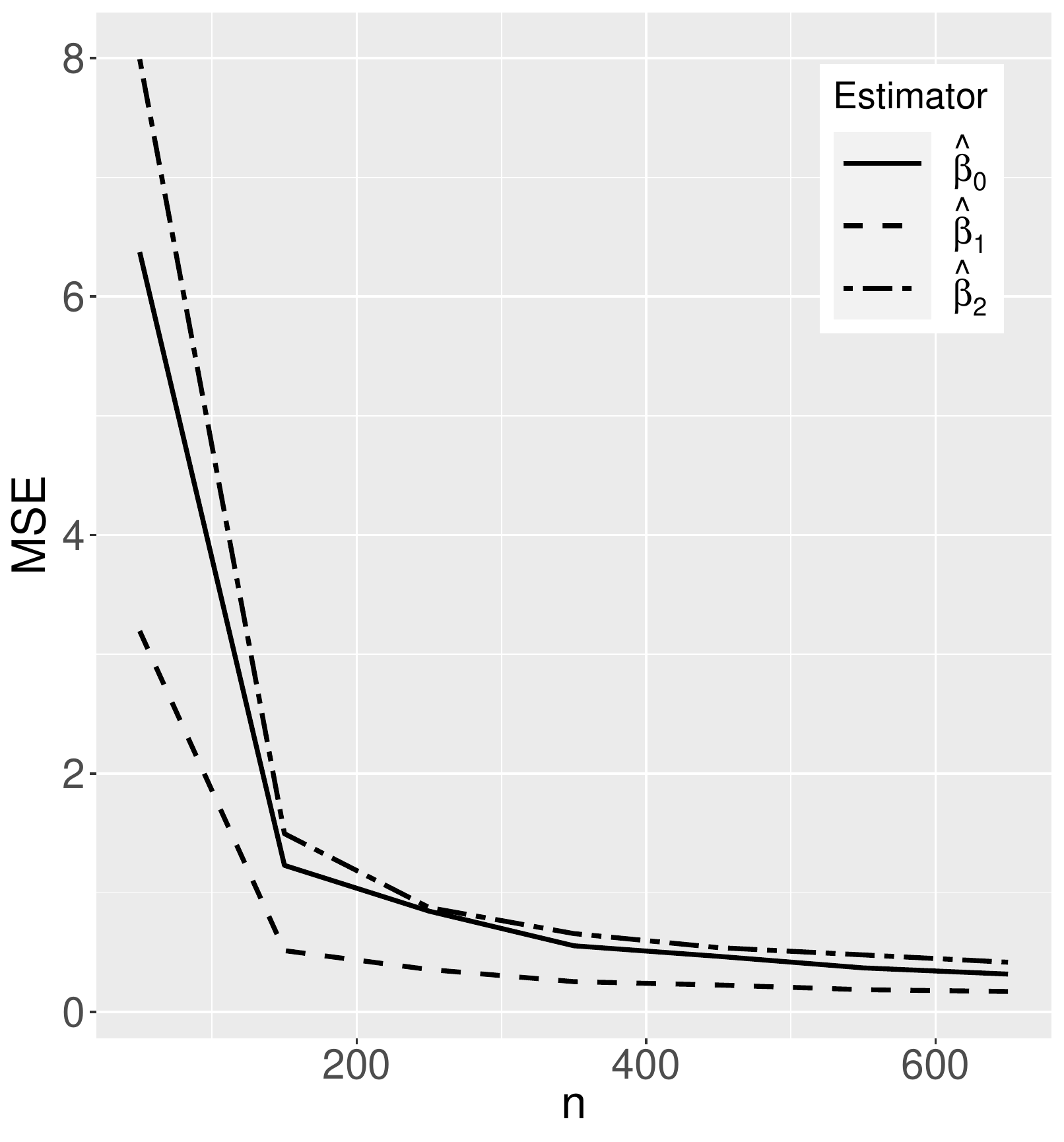}}
\subfigure[$\widehat{\textrm{Bias}}$($\widehat{\kappa}_i$)]{\includegraphics[height=5cm,width=5cm]{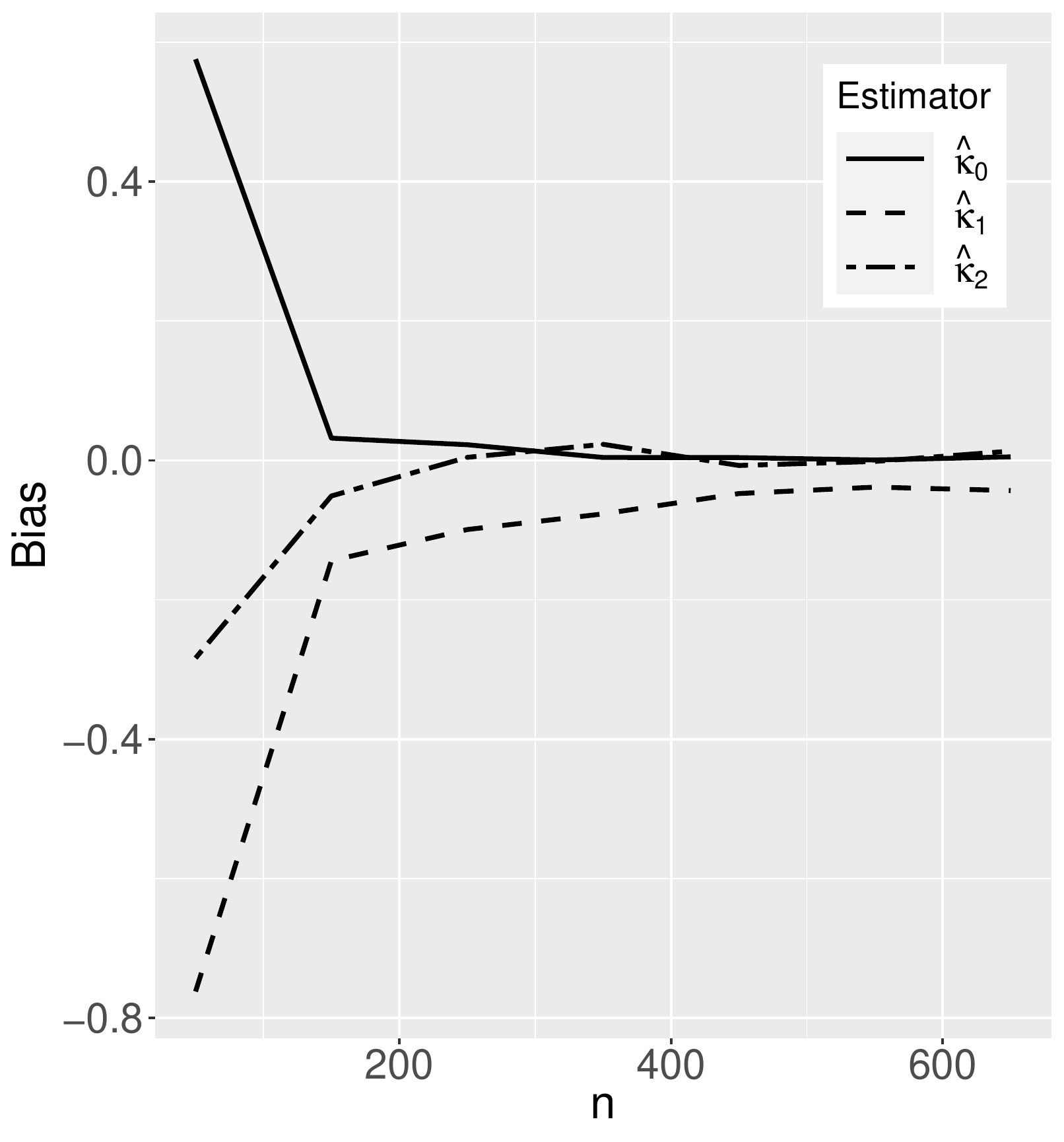}}
\subfigure[$\widehat{\textrm{MSE}}$($\widehat{\kappa}_i$)]{\includegraphics[height=5cm,width=5cm]{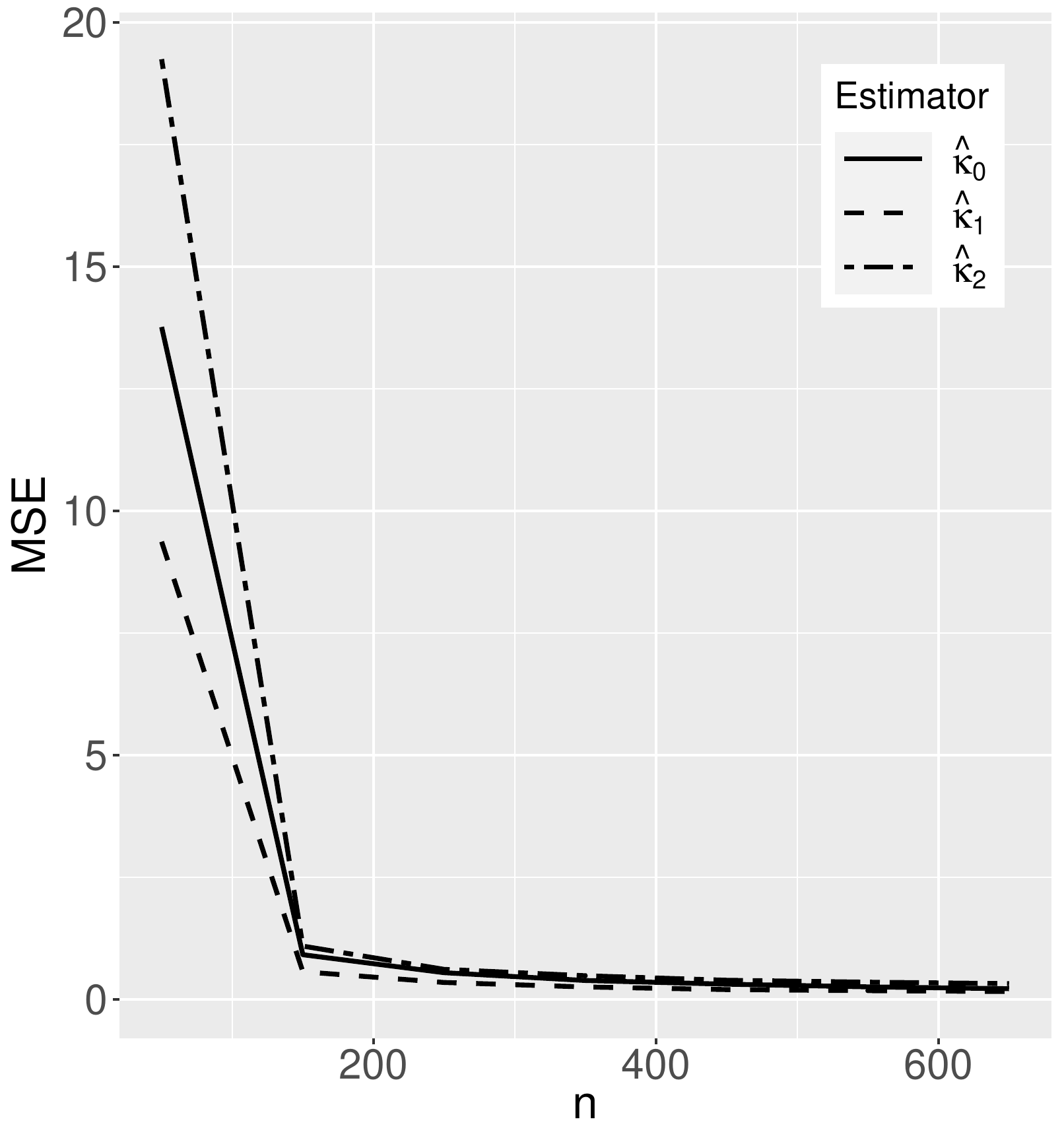}}
\subfigure[$\widehat{\textrm{Bias}}$($\widehat{\eta}_i$)]{\includegraphics[height=5cm,width=5cm]{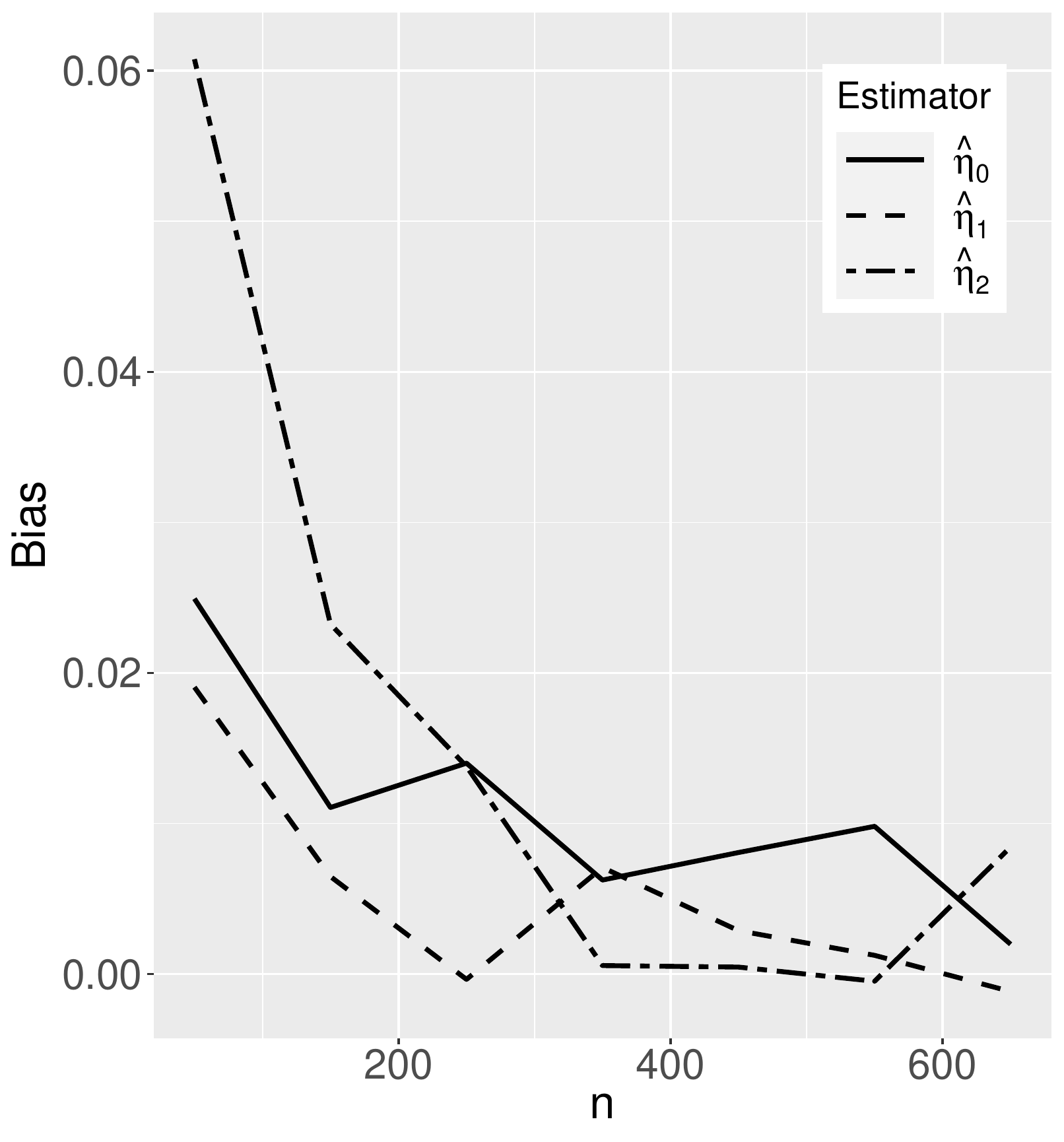}}
\subfigure[$\widehat{\textrm{MSE}}$($\widehat{\eta}_i$)]{\includegraphics[height=5cm,width=5cm]{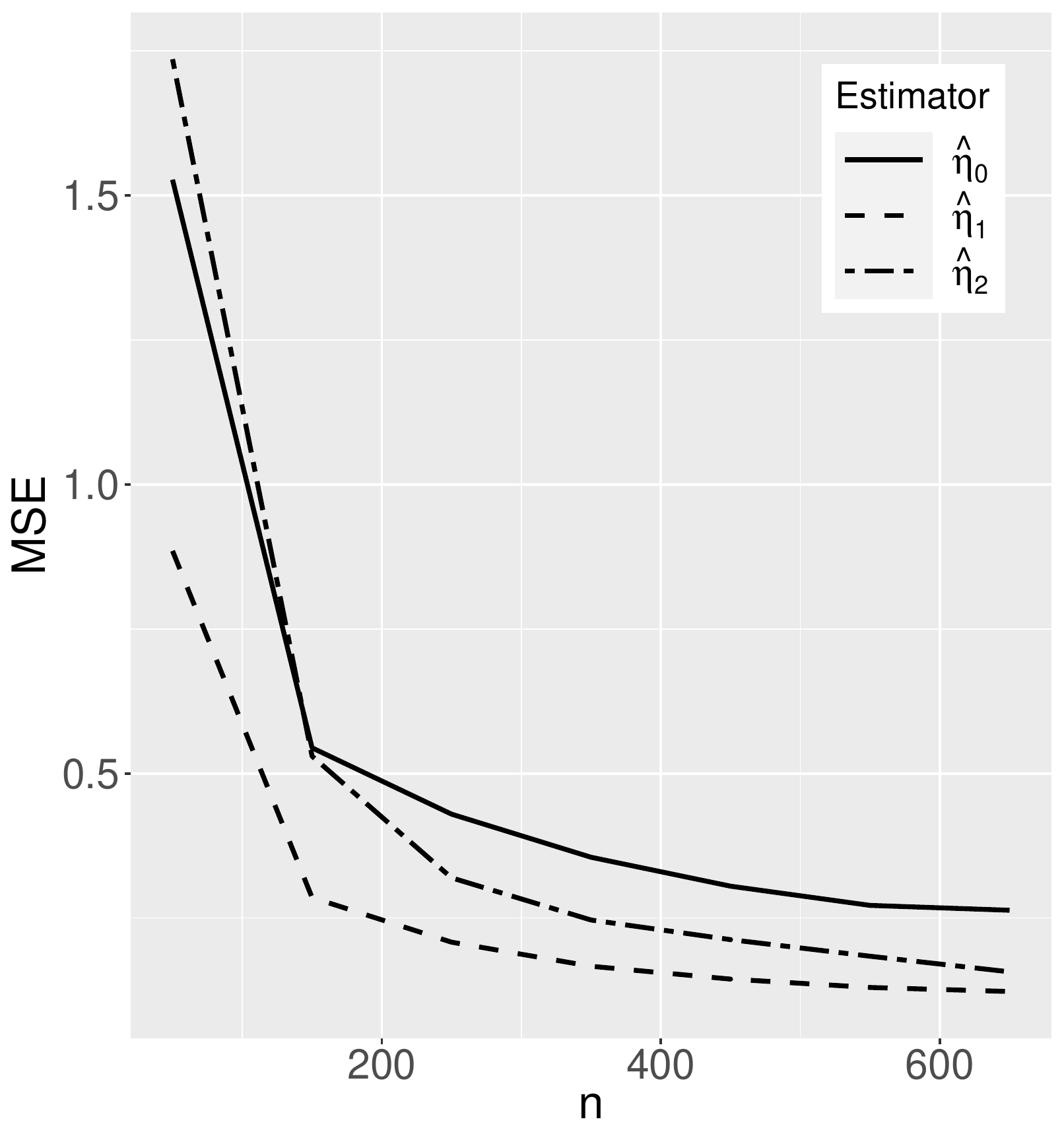}}
 \caption{\small {Bias and MSE estimates for $q=0.50$ ($i=\{0,1,2\}$).}}
\label{fig:sim2}
\end{figure}

\begin{figure}[!ht]
\centering
\subfigure[$\widehat{\textrm{Bias}}$($\widehat{\beta}_i$)]{\includegraphics[height=5cm,width=5cm]{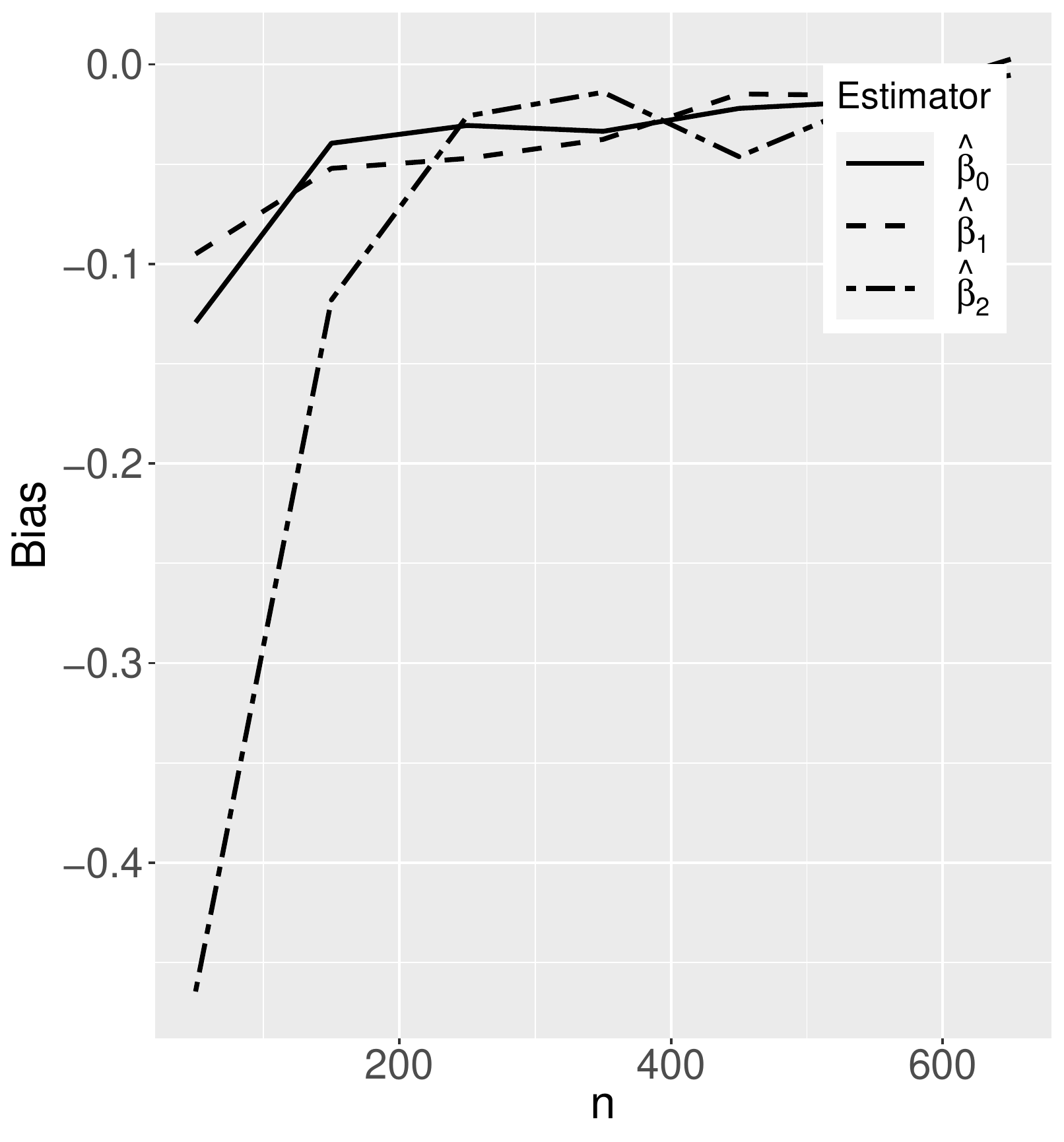}}
\subfigure[$\widehat{\textrm{MSE}}$($\widehat{\beta}_i$)]{\includegraphics[height=5cm,width=5cm]{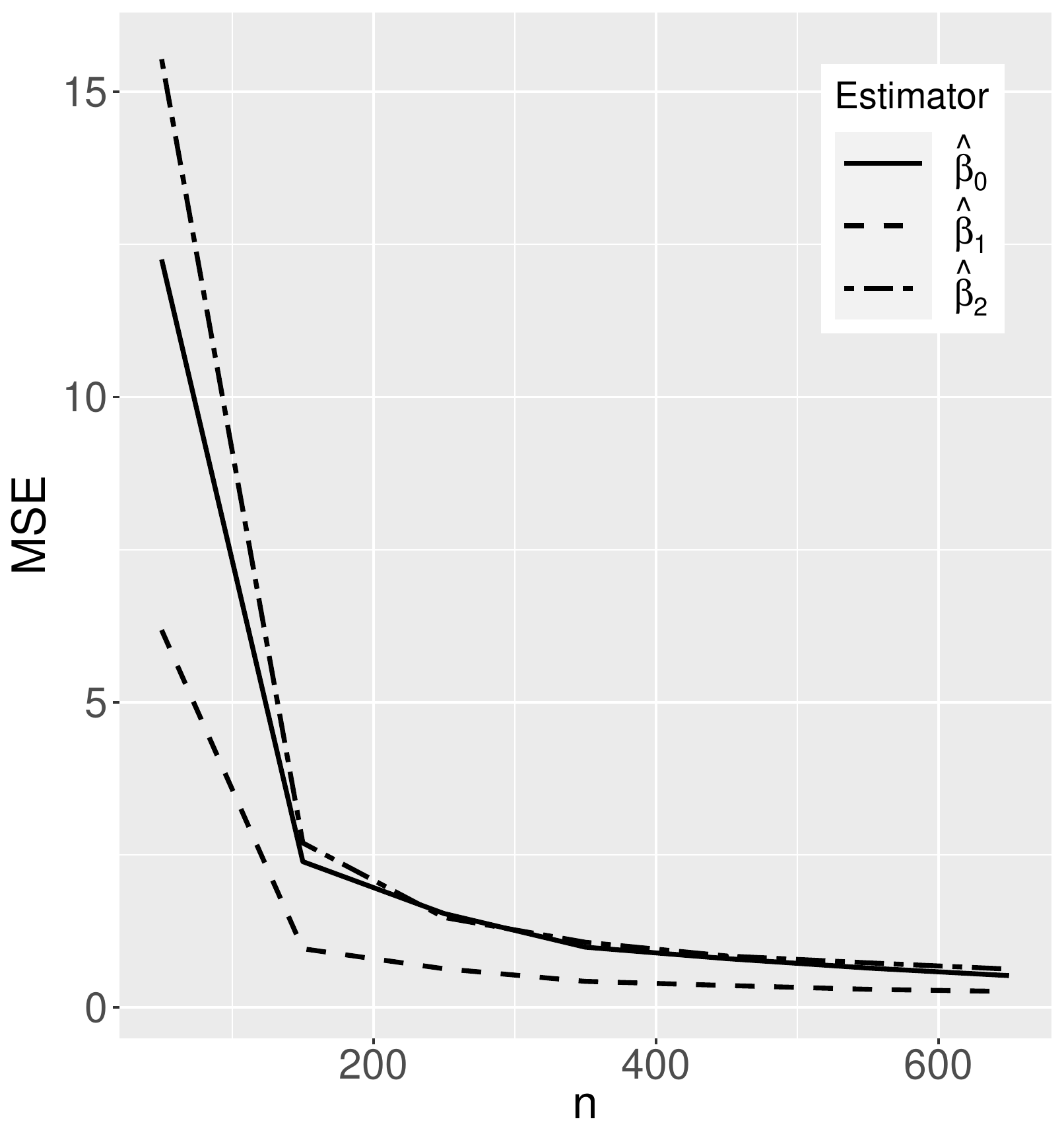}}
\subfigure[$\widehat{\textrm{Bias}}$($\widehat{\kappa}_i$)]{\includegraphics[height=5cm,width=5cm]{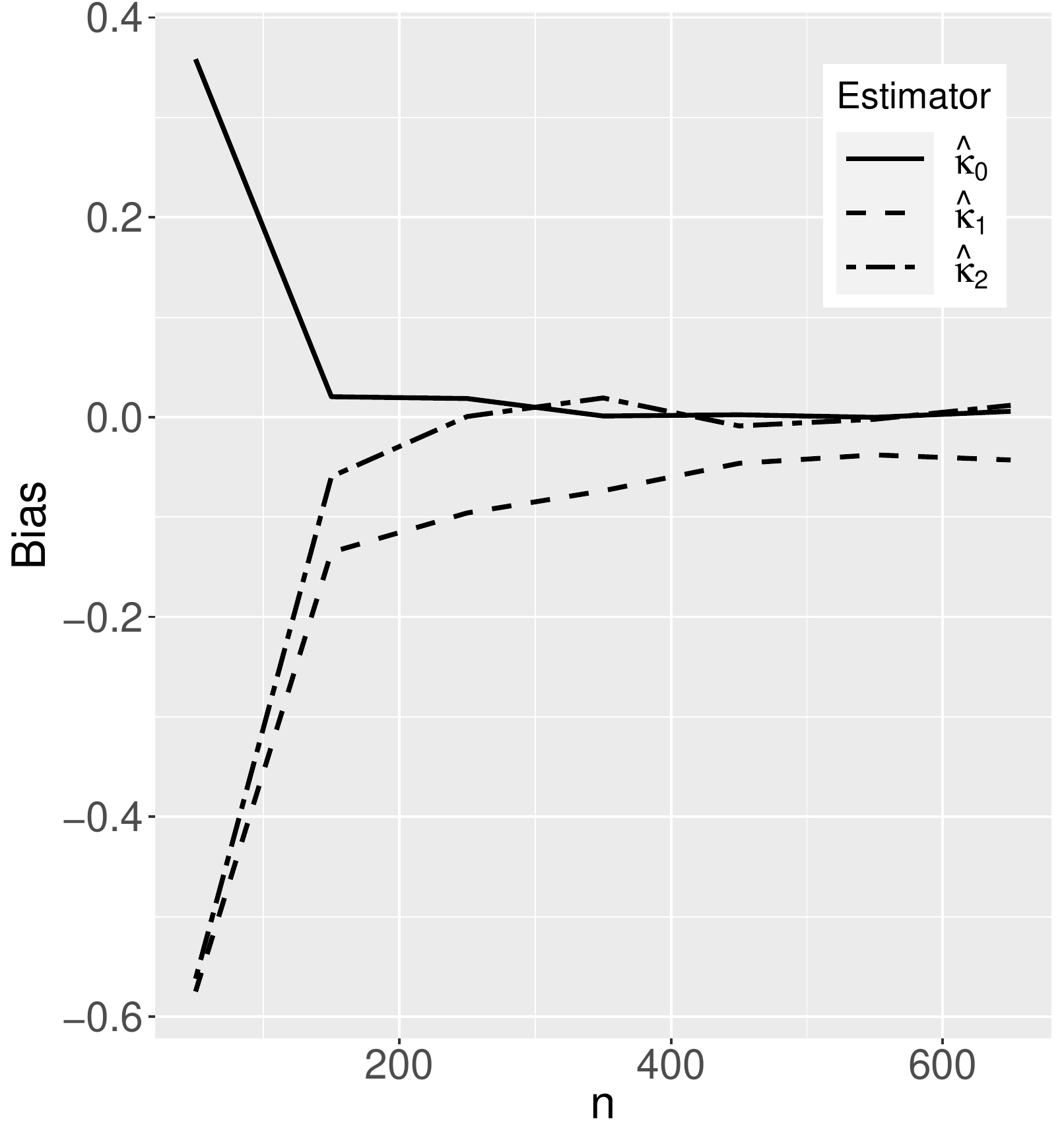}}
\subfigure[$\widehat{\textrm{MSE}}$($\widehat{\kappa}_i$)]{\includegraphics[height=5cm,width=5cm]{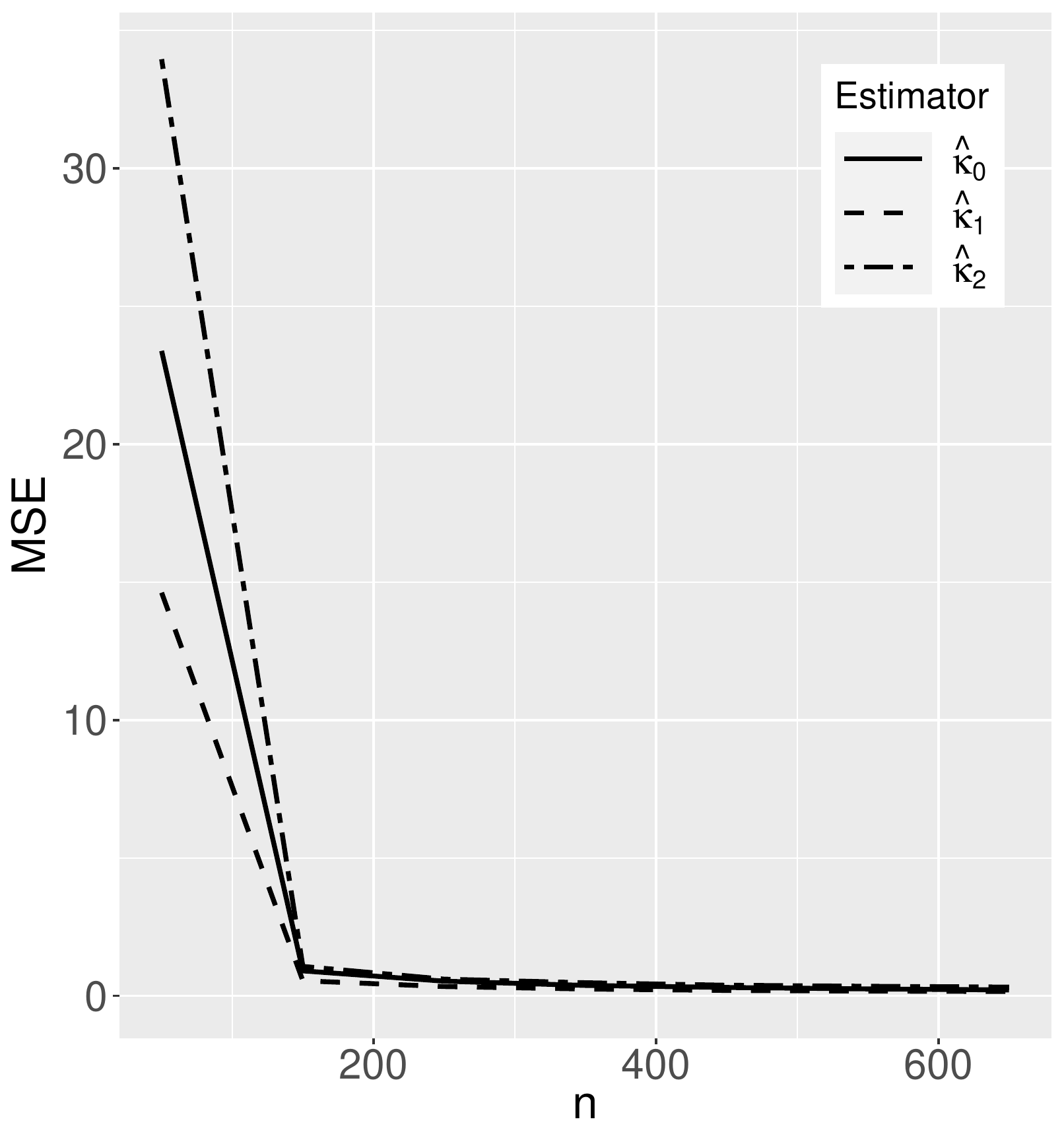}}
\subfigure[$\widehat{\textrm{Bias}}$($\widehat{\eta}_i$)]{\includegraphics[height=5cm,width=5cm]{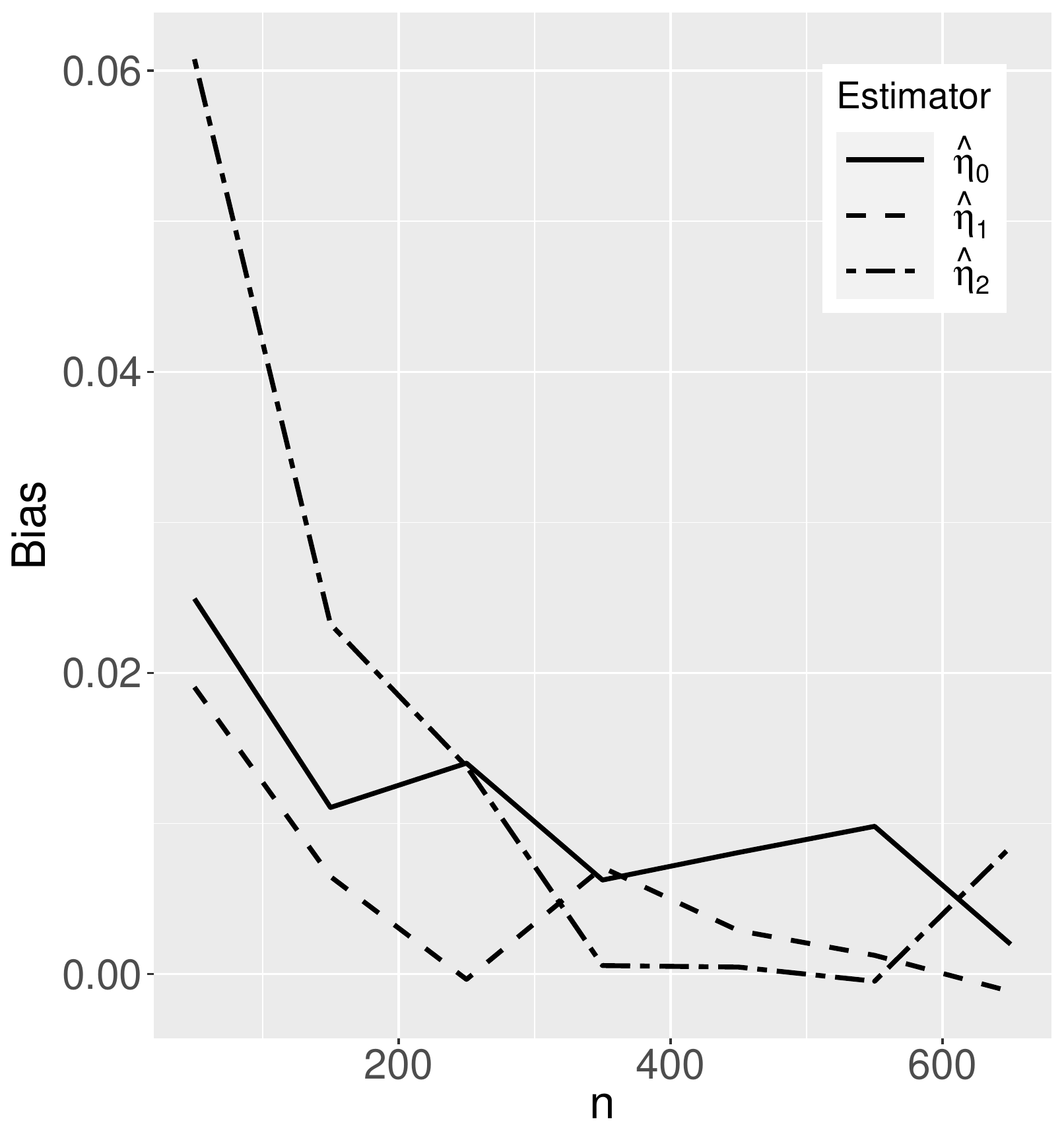}}
\subfigure[$\widehat{\textrm{MSE}}$($\widehat{\eta}_i$)]{\includegraphics[height=5cm,width=5cm]{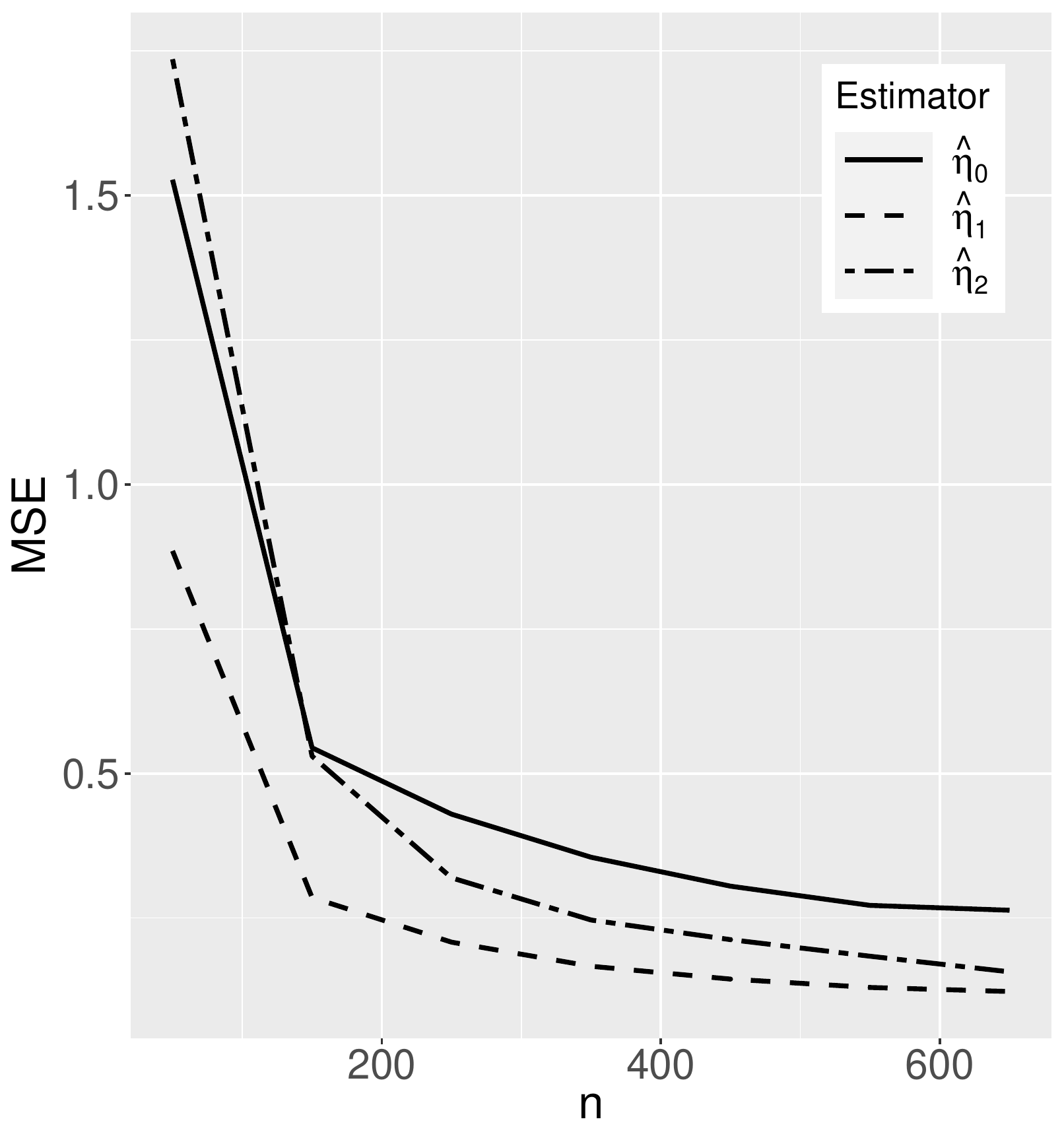}}
 \caption{\small {Bias and MSE estimates for $q=0.90$ ($i=\{0,1,2\}$).}}
\label{fig:sim3}
\end{figure}

A Monte Carlo simulation study is carried out to evaluate the performance of the maximum likelihood estimates of the \textit{quantile}-ZALS regression model. The performance is assessed using bias and mean square error (MSE) estimates, given by
$$\widehat{\textrm{Bias}}(\widehat{\theta}) = \frac{1}{\text{NREP}} \sum_{i = 1}^{\text{NREP}} \widehat{\theta}^{(i)} - \theta \quad
\text{and} \quad
\widehat{\mathrm{EQM}}(\widehat{\theta}) = \frac{1}{\text{NREP}} \sum_{i = 1}^{\text{NREP}} (\widehat{\theta}^{(i)} - \theta)^2,$$
where $\theta$ and $\widehat{\theta}^{(i)}$ denote the true parameter value and its respective $i$-th maximum likelihood estimate, respectively, and $\text{NREP}$ is the number of Monte Carlo replicas. We use the \texttt{R} software has been used to do all numerical calculations; see \cite{r2020vienna}.
The model used to generate the samples is given by
\[
Q_i = \exp \left(\beta_0 + \beta_{1}x_{i1} + \beta_{2}x_{i2} \right), \,\, \phi_i = \exp \left(\kappa_0 + \kappa_{1}w_{i1} + \kappa_{2}w_{i2} \right); \text{and} \;  \pi_i = \frac{\exp \left(\eta_0 + \eta_{1}v_{i1} + \eta_{2}v_{i2}\right)}{1+\exp \left(\eta_0 + \eta_{1}v_{i1} + \eta_{2}v_{i2}\right)}, 
\]
where the reference distribution is log-normal (the results of the log-Student-$t$, log-power-exponential and extended Birnbaum-Saunders are similar and are therefore omitted here). The simulation has the following setting: $(\beta_0,\beta_1,\beta_2)= (0.5,0.7,1.0)$, $(\kappa_0,\kappa_1,\kappa_2)= (0.5,0.8,1.0)$, $(\eta_0,\eta_1,\eta_2)= (0.5,0.3,0.5)$, 
with $\text{NREP}=5,000$ Monte Carlo replicas.The explanatory variables $x_i$, $w_i$ and $v_i$ are generated from the Uniform(0,1) distribution. Figures \ref{fig:sim1}-\ref{fig:sim3} show the Monte Carlo simulation results for $q=\{0.10,0.50,0.90\}$. An analysis of the results allows us to conclude that, in general, as the sample size increases, the bias and MSE decrease, as expected.

\section{Application to extramarital affairs data}\label{sec:5}

In this subsection, the \textit{quantile}-ZALS regression models are illustrated using data from extramarital affairs. The database is available at \cite{fair:78} and the objective here is to study the allocation of time in extramarital affairs for men and women married for the first time. There are 6,366 observations in the database, where the dependent variable ($Z$) is the time spent in extramarital affairs, and the explanatory variables are
\begin{itemize}
\item \texttt{ratemarr}: rating of the marriage, coded 1 to 4;
\item \texttt{age}: age, in years;
\item \texttt{yrsmarr}:  number of years married;
\item \texttt{numkids}: number of children;
\item \texttt{relig}: religiosity, code 1 to 4, 1 = not, 4 = very;
\item \texttt{educ}: education, coded 9, 12, 14, 16, 17, 20, that is, 9 = elementary school, 12 = high school, . . . , 20 = doctorate or other;
\item \texttt{wifeocc}: wife's occupation - Hollingshead scale;
\item \texttt{husocc}: husband's occupation - Hollingshead scale.
\end{itemize}

\begin{figure}[!ht]
\centering
\subfigure[]{\includegraphics[height=5.5cm,width=5.5cm]{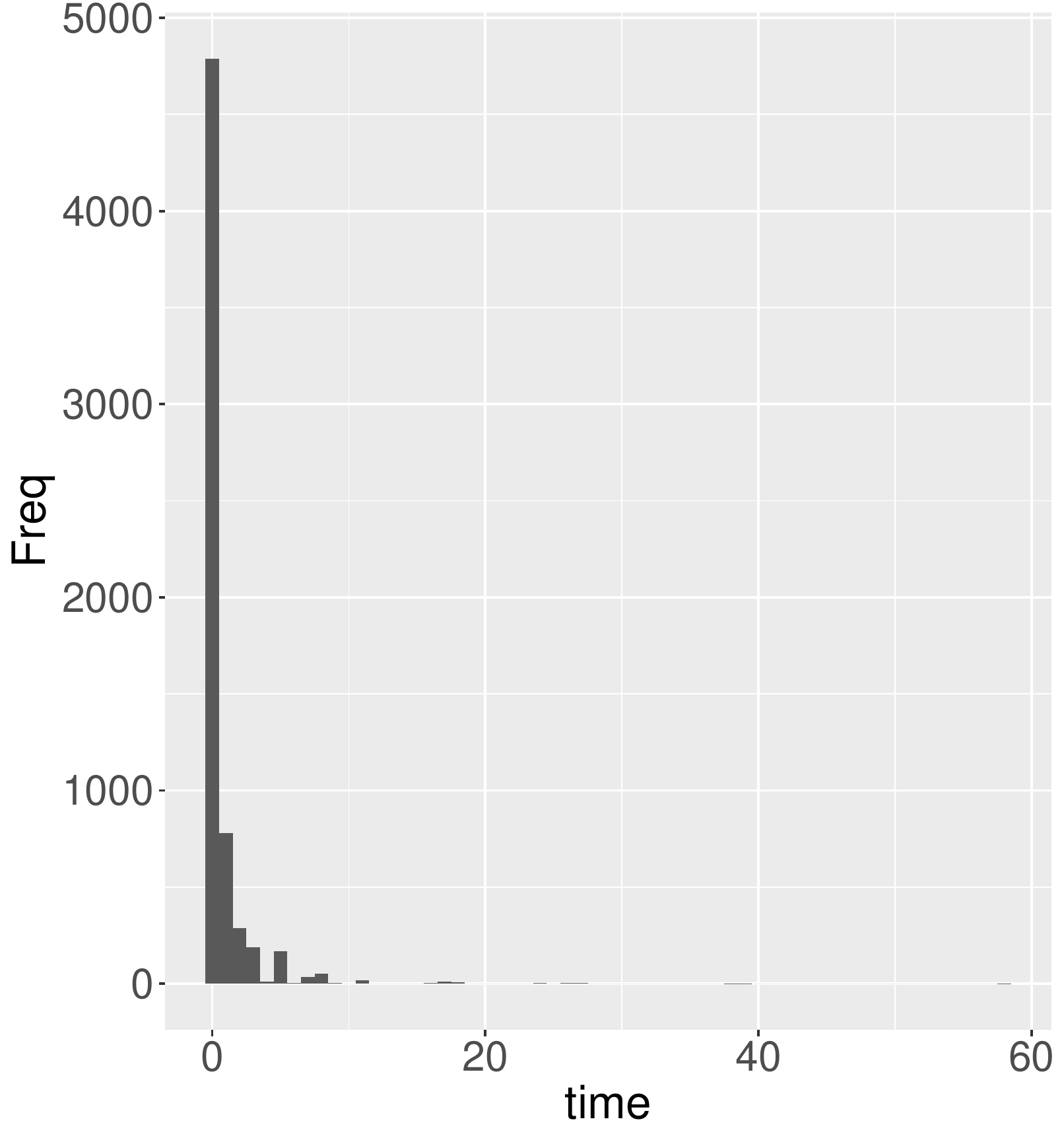}}
\subfigure[]{\includegraphics[height=5.5cm,width=5.5cm]{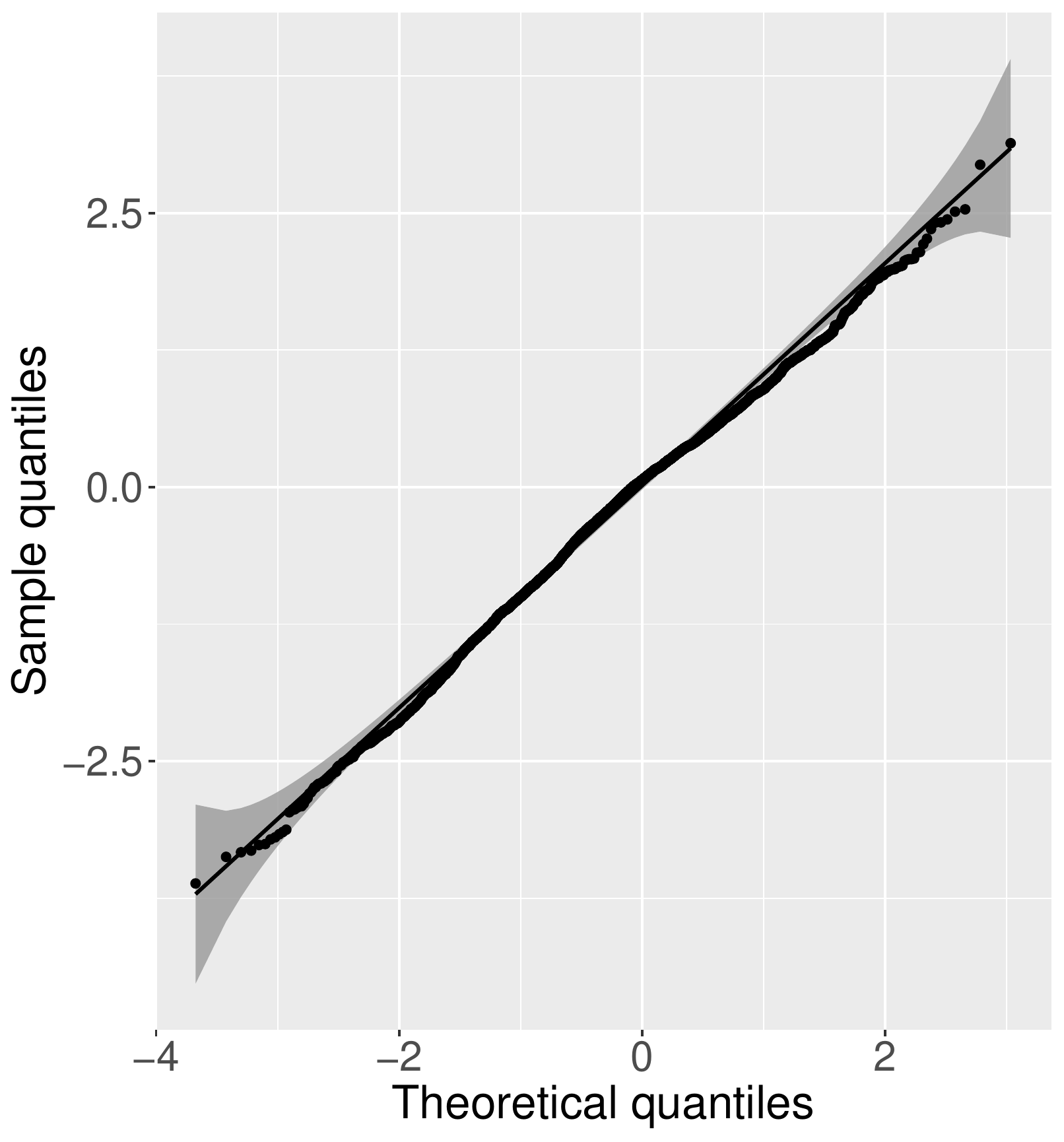}}
 \caption{\small {Histogram (a) for the time spent in extramarital affairs and QQ plot (b) and its envelope for the randomized quantile residuals based on the \textit{quantile}-ZAEBS regression model ($q=0.50$).}}
\label{fig:histextra}
\end{figure}

Descriptive statistics for the time spent in extramarital affairs ($Z$) indicate that the mean, median and standard deviation are given by 0.705, 0 and 2.203, respectively. The coefficient of variation is 312.37\%, indicating a high dispersion of data around the mean. The coefficients of skewness and kurtosis are equal to 8.761 and 131.912, respectively, which shows the presence of a high positive skewness and the presence of heavy tails. Then, the hypothesis of the use of log-symmetric distributions is plausible. Note that the asymmetric nature of the data is confirmed by the histogram shown in Figure \ref{fig:histextra}(a). Note also a high concentration of zero values in the sample, about 4,313 individuals have 0 time spent in extramarital affairs.

The proposed \textit{quantile}-ZALS regression models can accommodate heteroscedasticity, then specifications with and without explanatory variables in the relative dispersion $\phi$ can be fitted. We consider here \textit{quantile}-ZALS regression models based on the log-normal (\textit{quantile}-ZALNO), log-Student-$t$ (\textit{quantile}-ZAL$t$), log-power-exponential (\textit{quantile}-ZALPE) and extended Birnbaum-Saunders (\textit{quantile}-ZAEBS) distributions. An analysis of the significance of the coefficients suggested the following specification:

\clearpage 
\begin{eqnarray*}\label{fsdfds}\nonumber
&Q_i    = \exp\left(
\beta_0 + \beta_1 \texttt{ratemarr}  + \beta_2\texttt{yrsmarr} + \beta_3\texttt{numkids} + \beta_4\texttt{relig} \right) \\ \nonumber
&\phi_i = \exp\left( \kappa_0 +  \kappa_1  \texttt{age}  \right) \,\, \mbox{and} \\ 
&\pi_i  = \Lambda\left(
\eta_0 + \eta_1 \texttt{ratemarr} + \eta_2  \texttt{age} + \eta_3\texttt{yrsmarr}  + \eta_4\texttt{relig} + \eta_5\texttt{educ} \right. \\
&  \left. + \eta_6\texttt{wifeocc} \right).
\end{eqnarray*}

Table \ref{tab:resultssaicbicall} reports the results of the averages of the AIC and BIC values based on $q=\{0.01,0.02,\ldots,0.99\}$ for the proposed \textit{quantile}-ZALS regression models. The results indicate that the lowest values of AIC and BIC are those based on the \textit{quantile}-ZALPE and \textit{quantile}-ZAEBS models, with a slight advantage of the latter.

\begin{table}[ht]
\footnotesize
\centering
 \caption{\small {Averages of the of the AIC and BIC values with $q=\{0.01,0.02,\ldots,0.99\}$ for different models.}}
\begin{tabular}{lrrrrrrrrrrrrrr}
  \hline
                                        &  \multicolumn{4}{c}{Model}  \\ \cline{2-5}
                              Criterion  &  \textit{quantile}-ZALNO  & \textit{quantile}-ZAL$t$ & \textit{quantile}-ZALPE  &  \textit{quantile}-ZAEBS  \\ 
					\hline
                             AIC       & 13165.14 & 13173.48 & 13163.49 & 13163.41  \\ 
                             BIC       & 13259.77 & 13268.10 & 13258.12 & 13258.03  \\    
  \hline         
\end{tabular}
\label{tab:resultssaicbicall}
\end{table}

Since the results of the \textit{quantile}-ZAEBS model showed the best results, we can compare them with the results of the zero-adjusted gamma (ZAGA) and zero-adjusted inverse Gaussian (ZAIG) regression models, studied by \cite{Heller2006} and \cite{stasinopoulosetal:17}. Table \label{tab:aicompar} reports the results of AIC and BIC for these models, and we observe that the \textit{quantile}-ZAEBS model has the best fit. The QQ plot for the randomized quantile residuals \citep{dunnsmyth:96} of this model for $q = 0.50$ is shown in Figure \ref{fig:histextra}(b); similar plots are obtained for other values of $q$. The results then indicate that the proposed model present adjustments that are superior to existing models in the literature.

\begin{table}[ht]
\footnotesize
\centering
 \caption{\small {AIC and BIC values for different \textit{quantile}-ZALS regression models.}}
\begin{tabular}{lrrrrrrrrrrrrrr}
  \hline
           &           &  \multicolumn{3}{c}{Model}  \\ \cline{3-5}
           &           &  \textit{quantile}-ZAEBS  & ZAGA & ZAIG \\ 
           &           &  ($q=0.50$)  &   &   \\ 
					\hline
AIC        &           & 13163.18 & 13466.62 & 13953.84 \\ 
BIC        &           & 13257.80 & 13580.48 & 14067.71   \\
  \hline         
\end{tabular}
\label{tab:aicompar}
\end{table}

\begin{figure}[!ht]
\centering
\subfigure[$\beta_0$]{\includegraphics[height=5cm,width=5cm]{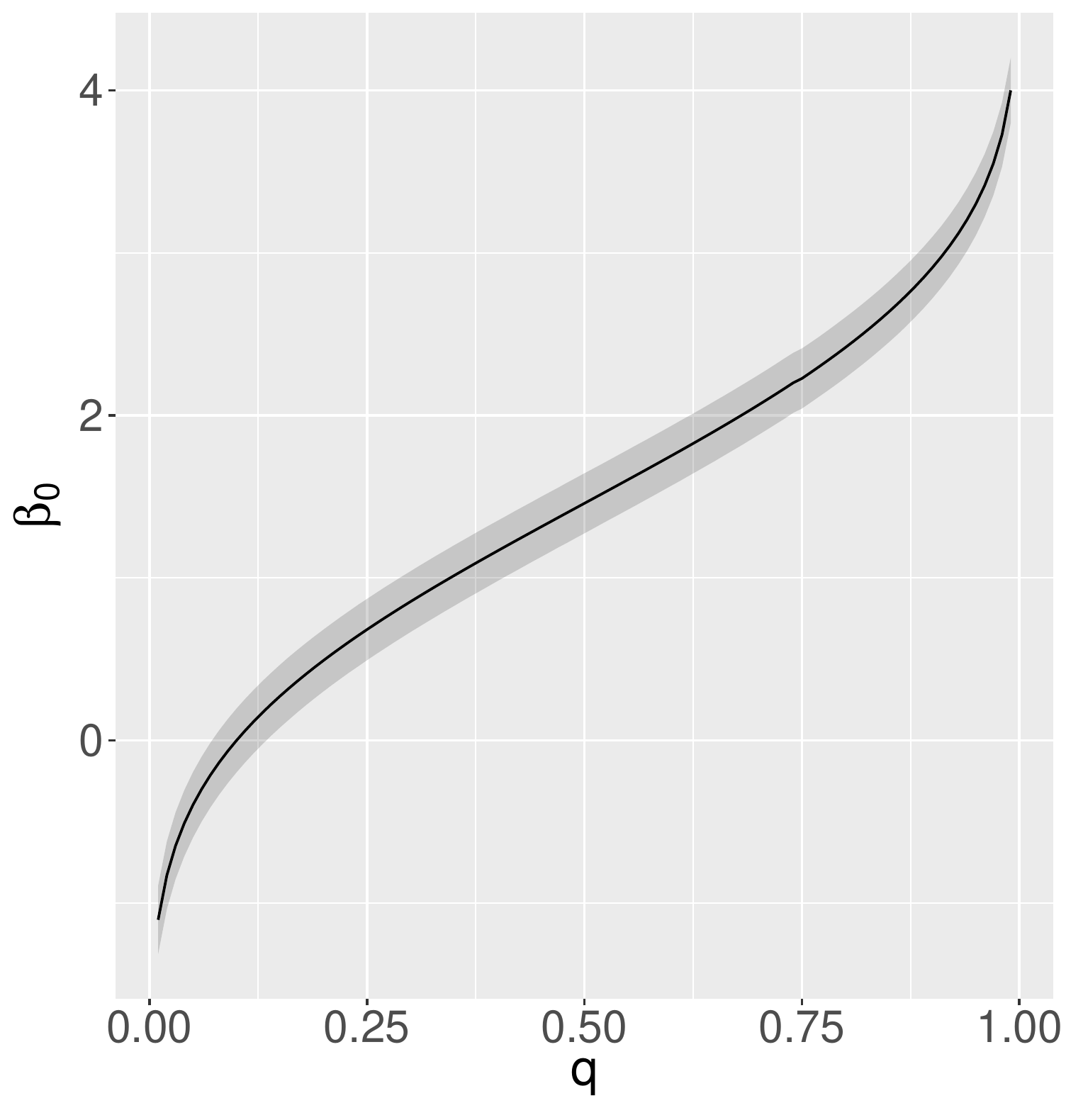}}
\subfigure[$\beta_1$]{\includegraphics[height=5cm,width=5cm]{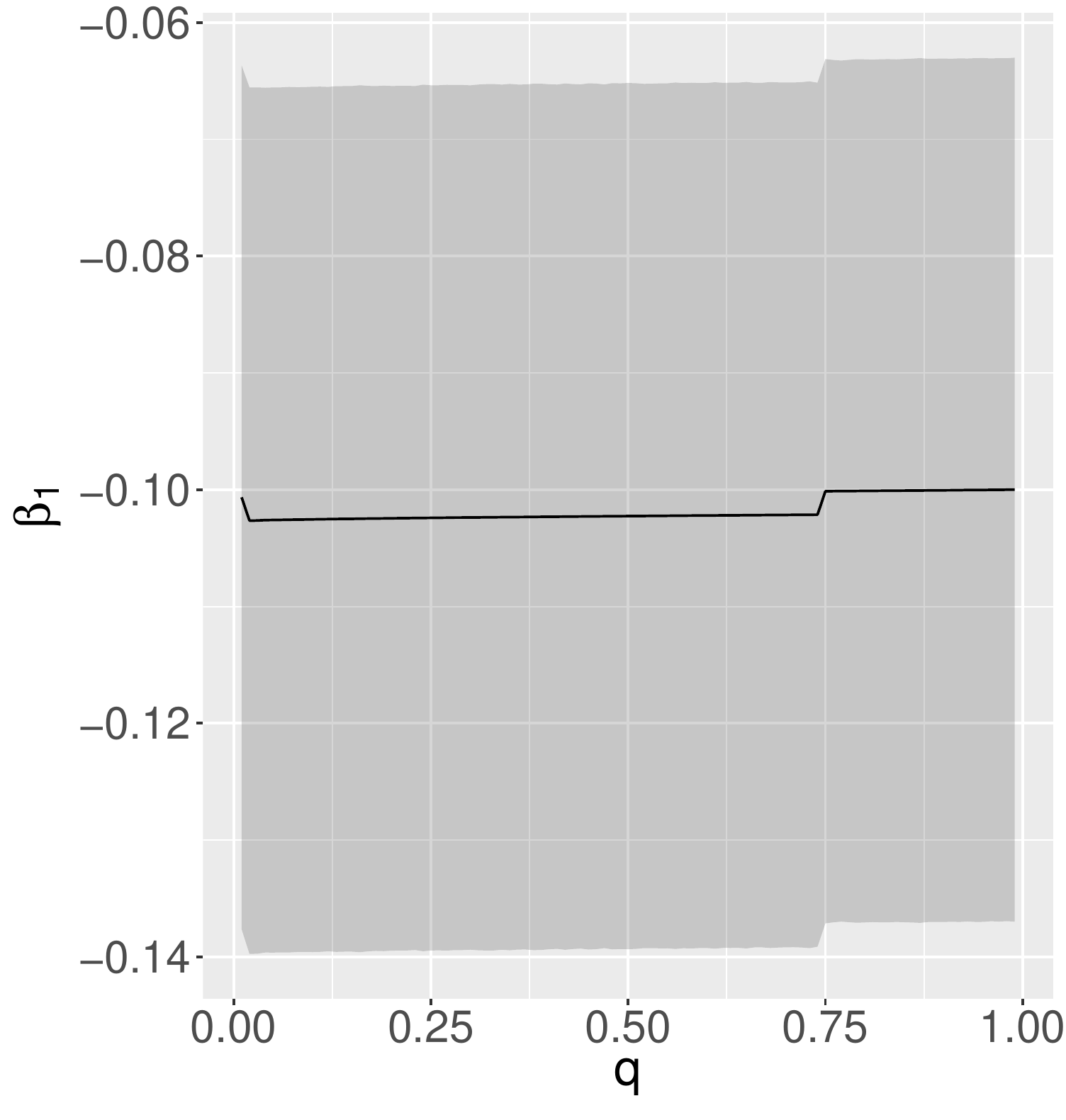}}
\subfigure[$\beta_2$]{\includegraphics[height=5cm,width=5cm]{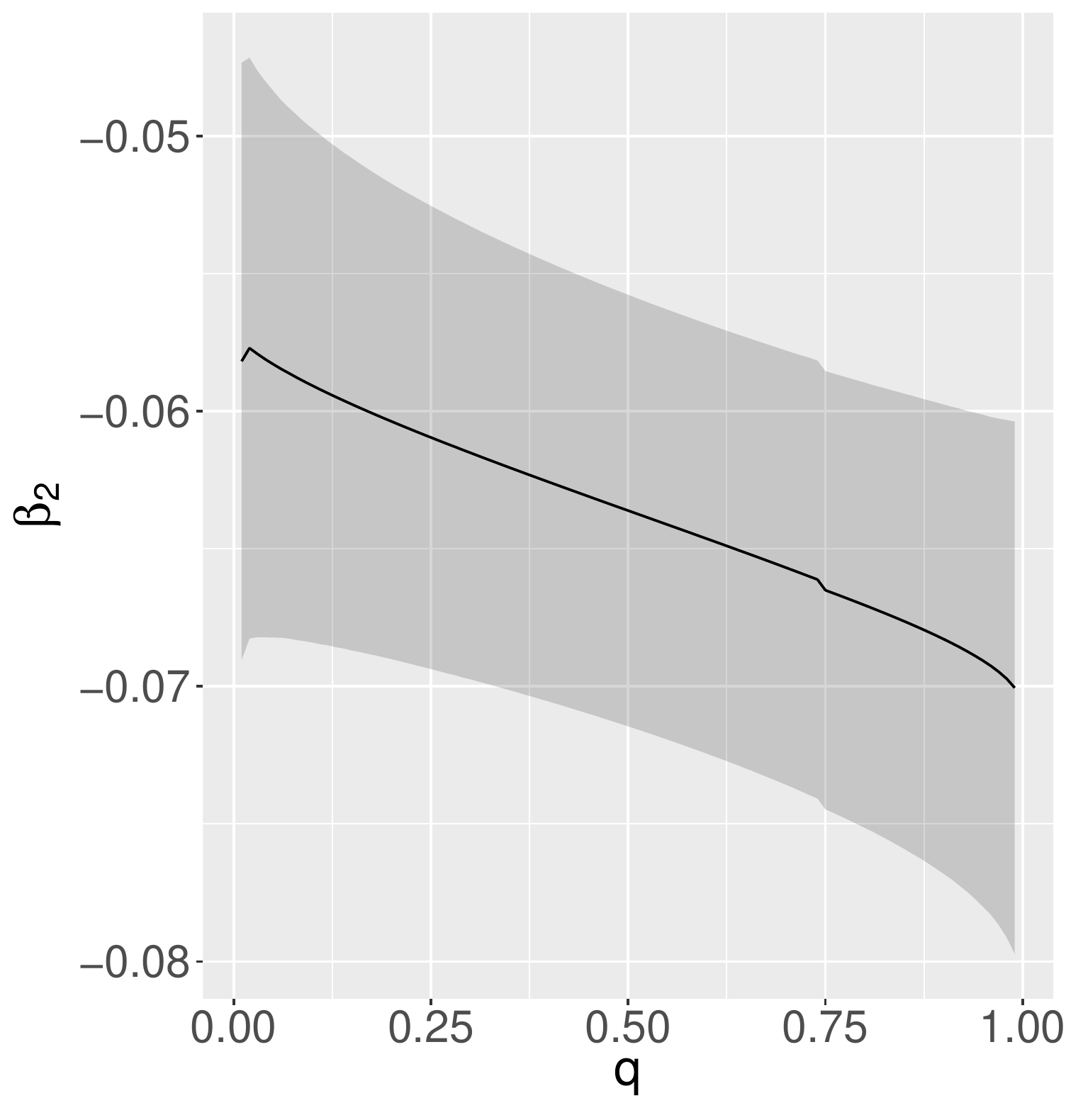}}\\
\subfigure[$\beta_3$]{\includegraphics[height=5cm,width=5cm]{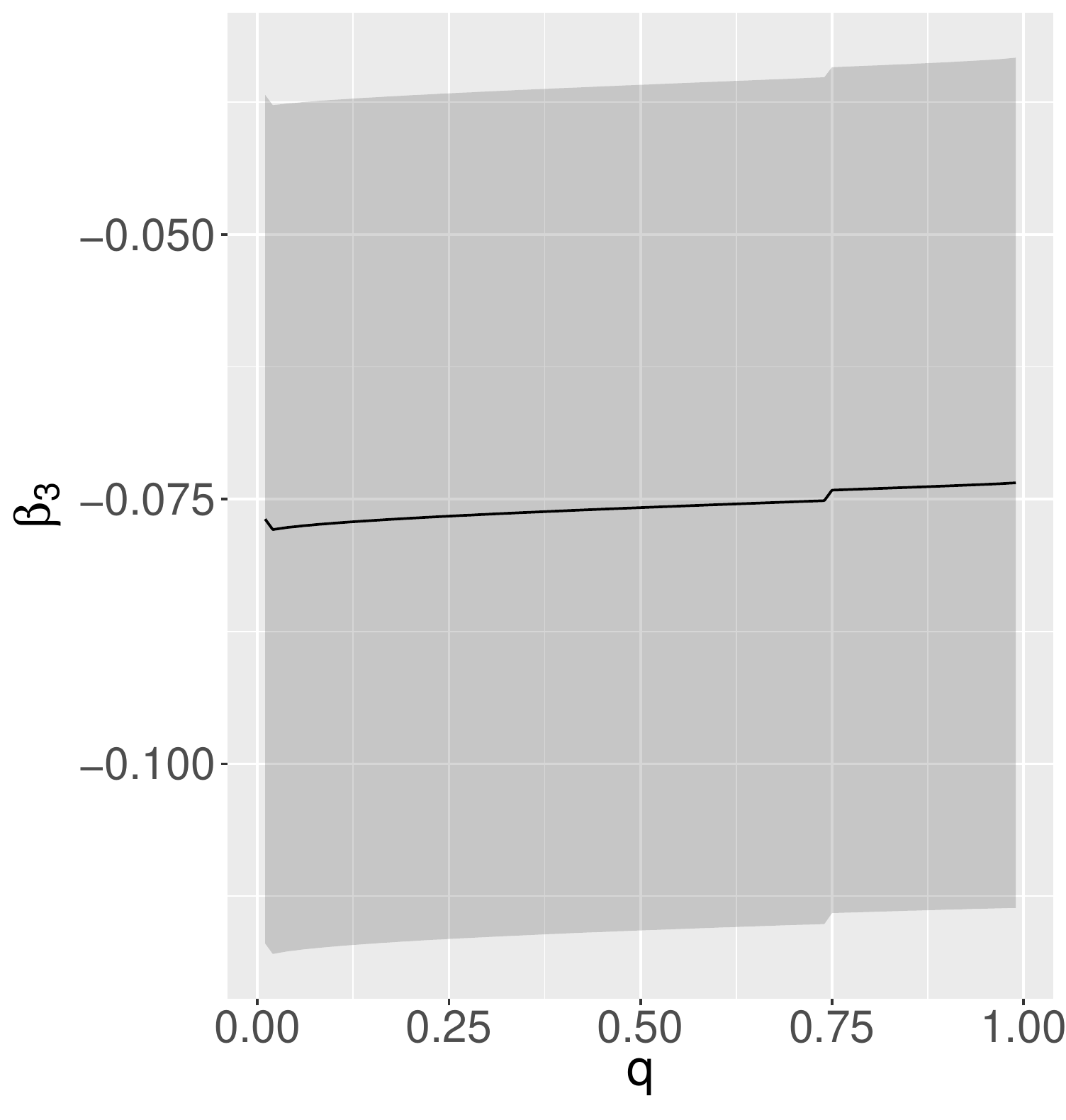}}
\subfigure[$\beta_4$]{\includegraphics[height=5cm,width=5cm]{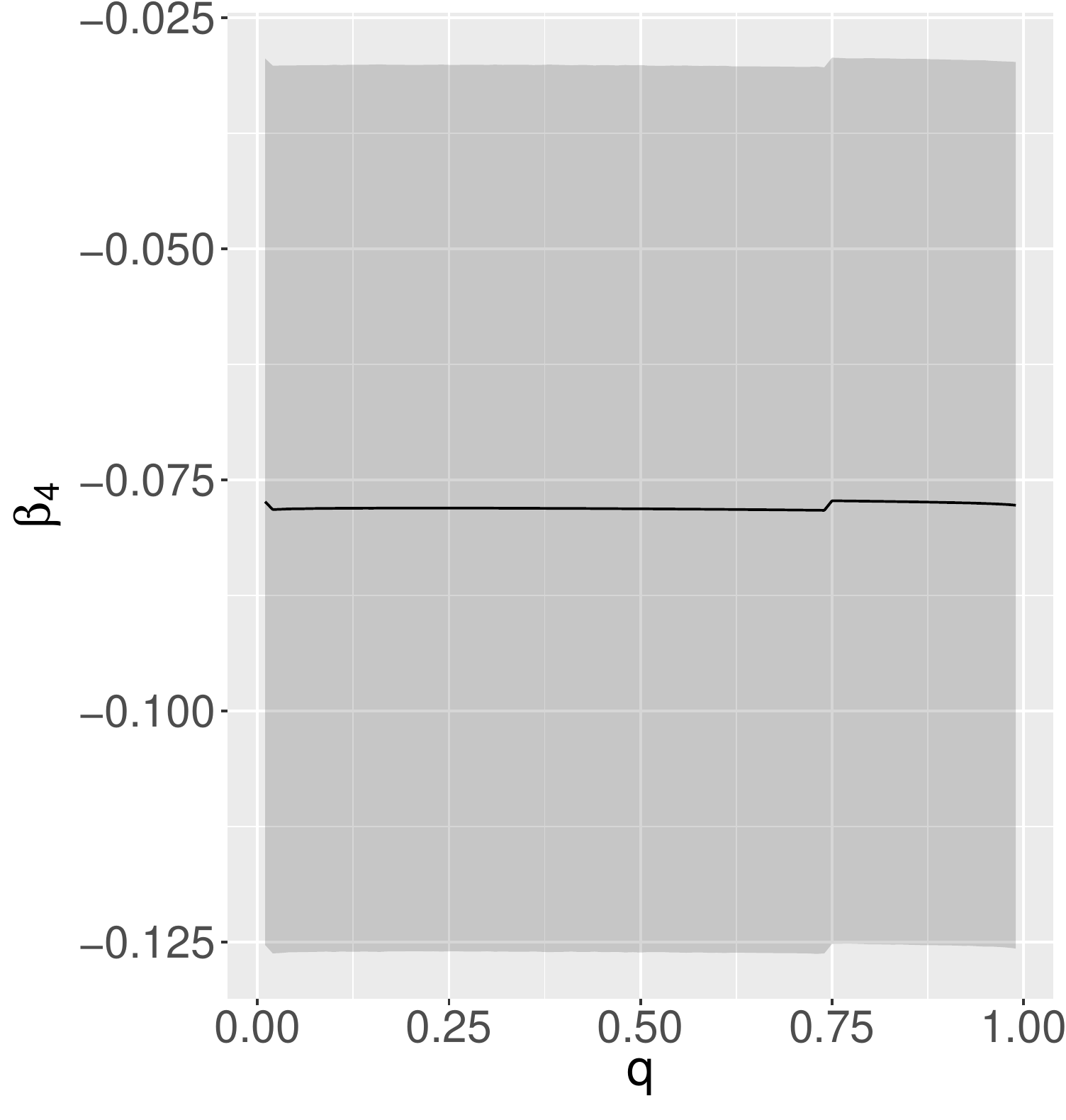}}
\subfigure[$\kappa_0$]{\includegraphics[height=5cm,width=5cm]{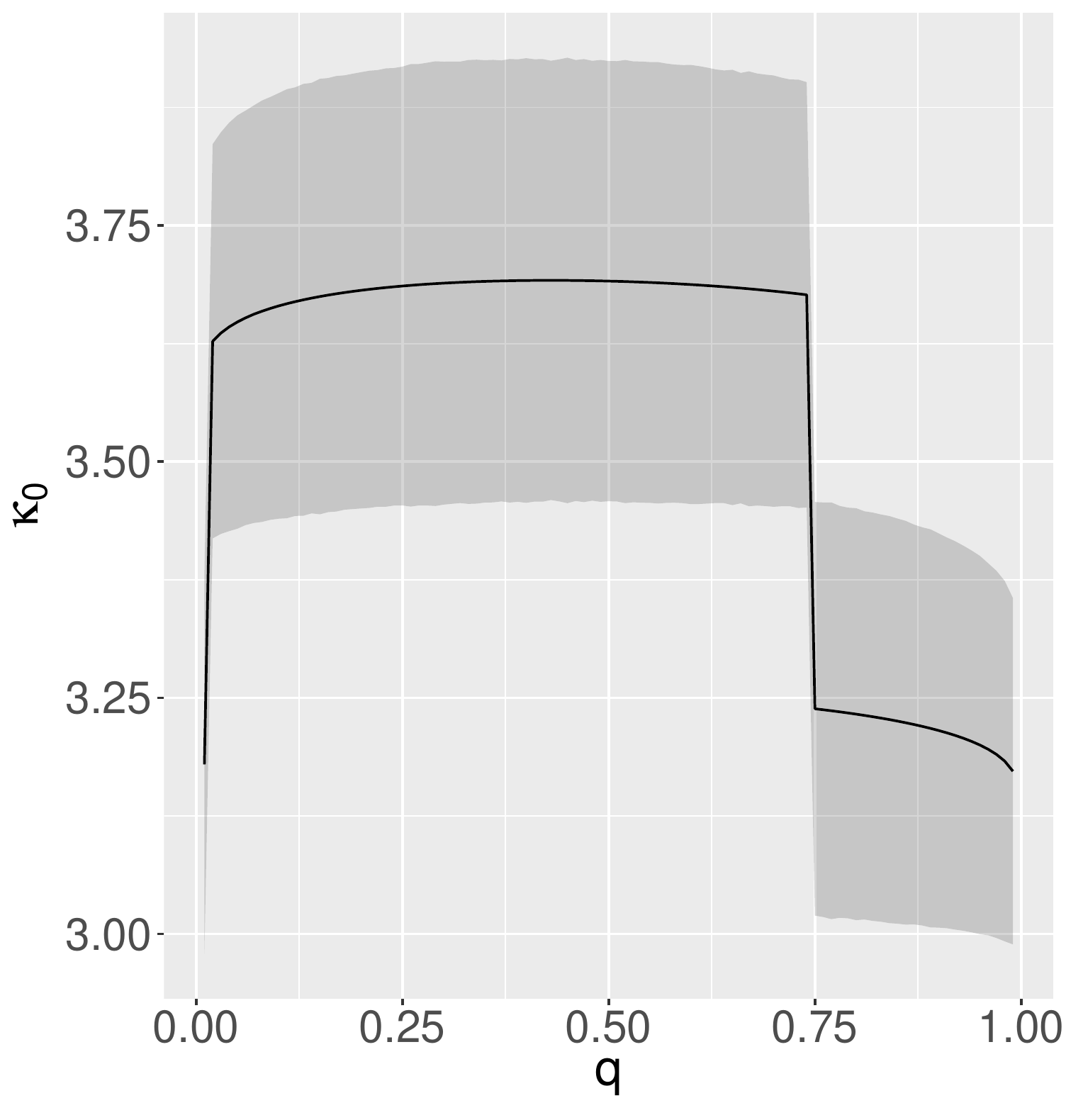}}\\
\subfigure[$\kappa_1$]{\includegraphics[height=5cm,width=5cm]{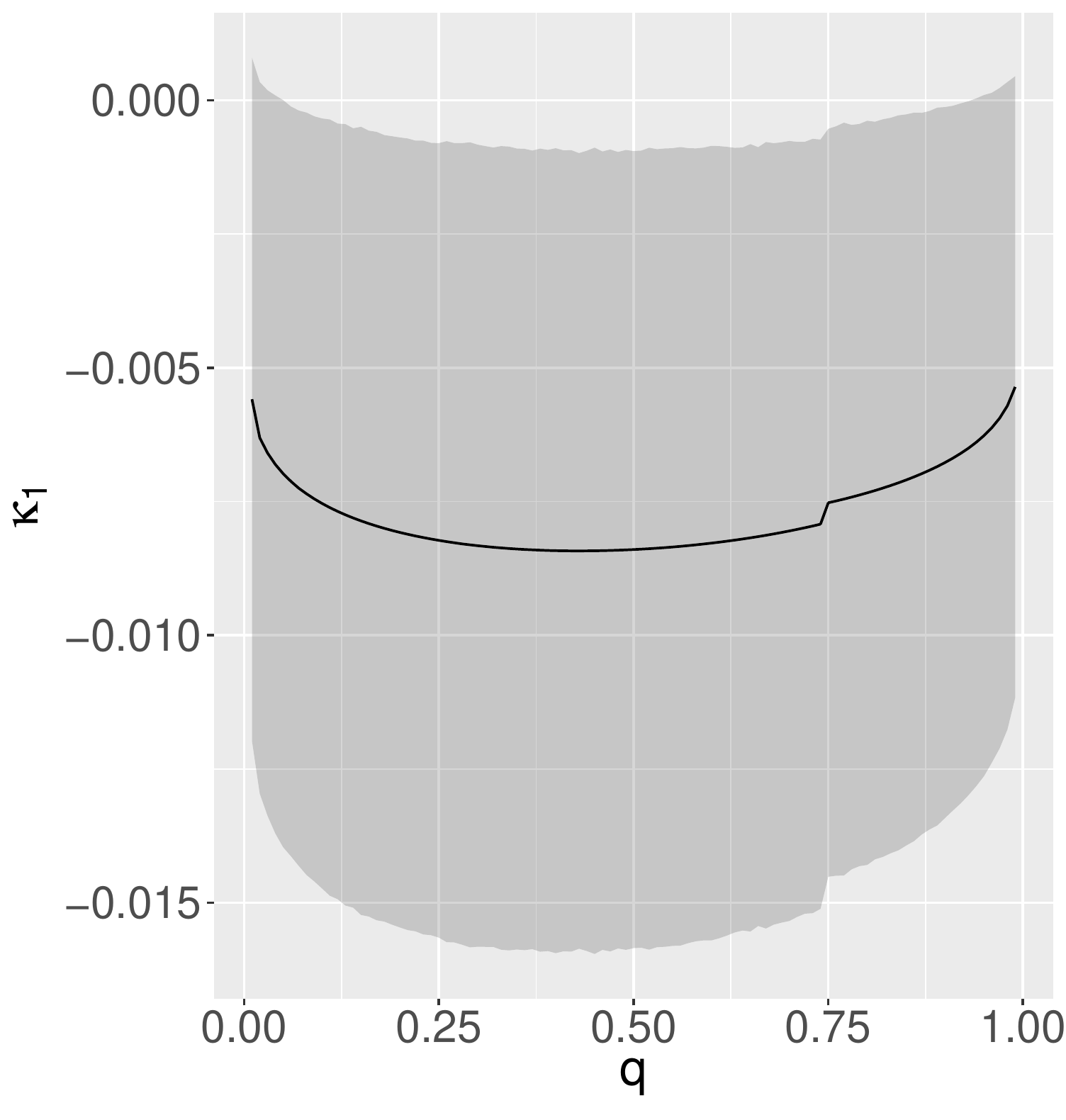}}
 \caption{\small {Parameter estimates (confidence intervals in grey) for the positive continuous part of the \textit{quantile}-ZAEBS regression model.}}
\label{fig:estimates}
\end{figure}

The estimates of the parameters of the \textit{quantile}-ZAEBS model for $\ pi_i$ (discrete component) are shown in Table \ref{tab:partdiscreta}, and those for $Q_i$ and $\phi_i$ in Figure \ref{fig:estimates}. The figure shows asymmetric dynamics. For example, the estimates of $\beta_0$ ($\beta_2$) tend to increase (decrease) with the increase of $q$. In general, such results show the importance of considering a quantile approach.

\begin{table}[!ht]
\footnotesize
\centering
 \caption{\small {Estimated parameters (standard errors in parentheses) of the discrete part of the \textit{quantile}-ZAEBS regression model.}}
\begin{tabular}{lrrrrrrrrrrrrrr}
  \hline
   &\texttt{Interc.}   & \texttt{ratemarr}&\texttt{age}&\texttt{yrsmarr}&\texttt{relig}&\texttt{educ}&\texttt{wifeocc} \\ 
  \hline
Estimate & 3.7371*  &-0.7153*  &-0.0602* & 0.1095*  &-0.3760* &-0.0380*  & 0.1628*   \\ 
Standard error& (0.2961) &(0.0314)  &(0.0103)  &(0.0097) &(0.0346) &(0.0153)  &(0.0337) \\   
\hline
        \multicolumn{8}{l}{\scriptsize{* significant at 5\% level. ** significant at 10\% level.}}
\end{tabular}
\label{tab:partdiscreta}
\end{table}

\section{Concluding remarks}\label{sec:6}

In this work, a class of zero-adjusted log-symmetric quantile regression models was proposed. The proposed regression model is based on a zero-adjusted version of the log-symmetric distributions parameterized by the quantile, also proposed in this work. The quantile proposal provides wide flexibility in the analysis of the effects of the explanatory variables on the dependent variable, which makes the proposed model a more interesting alternative than the existing zero-adjusted log-symmetric regression models \citep{cc:21,cosavalente:21}. The estimation of the parameters was performed by the maximum likelihood method and a Monte Carlo simulation study was carried out to evaluate the performance of the maximum likelihood estimates. An application of the proposed models is used to study the allocation of time in extramarital affairs of men and women married for the first time. The results showed that the proposed zero-adjusted log-symmetric quantile regression models perform better than the existing zero-adjusted gamma (ZAGA) and zero-adjusted inverse Gaussian (ZAIG) regression models studied by \cite{Heller2006} and \cite{stasinopoulosetal:17}.

\normalsize


\end{document}